\documentclass[epsf,useAMS,usenatbib,usegraphicx]{mn2e}
\usepackage{subfigure}
\usepackage{comment}
\usepackage{natbib}
\usepackage{graphicx}
\usepackage{amssymb}
\usepackage{amsbsy}   
\usepackage{amsmath}
\usepackage[ps2pdf,bookmarks=true,breaklinks=true,hypertexnames=false]{hyperref}
 \hypersetup{
pdfauthor = {Lee R. Spitler et al.},
pdftitle = {local epoch of reionization},
pdfsubject = {observational astrophysics},
pdfkeywords = {globular star clusters systems, galaxies, reionization},
pdfcreator = {LaTeX with hyperref package},
pdfproducer = {dvips + ps2pdf}}

\newcommand{\mh}{$M_{\rm vir}$~}
\newcommand{\msun}{$M_{\odot}$}

\newcommand{\mwnu}{$\nu=3.12\pm0.34$}
\newcommand{\mwnub}{$3.12\pm0.34$}
\newcommand{\mwnus}{$\nu=3.1\pm0.3$}
\newcommand{\mwnuss}{$\nu=3.1$}
\newcommand{\mwz}{$z_{reion}=12.1^{+1.6}_{-1.1}$}
\newcommand{\mwzplain}{$12.1^{+1.6}_{-1.1}$}

\newcommand{\nftosnub}{$2.86\pm0.36$}
\newcommand{\nftosnus}{$\nu=2.9\pm0.4$}
\newcommand{\nftosz}{$z_{reion}=11.0^{+1.7}_{-1.7}$}
\newcommand{\nftoszplain}{$11.0^{+1.7}_{-1.7}$}

\newcommand{\mesnu}{$\nu=2.19\pm0.41$}
\newcommand{\mesnub}{$2.19\pm0.41$}
\newcommand{\mesnus}{$\nu=2.2\pm0.4$}
\newcommand{\mesz}{$z_{reion}=7.8^{+2.0}_{-1.8}$}
\newcommand{\meszplain}{$7.8^{+2.0}_{-1.8}$}

\newcommand{\allnub}{$2.74\pm0.21$}

\newcommand{\allz}{$z_{reion}=10.5^{+1.0}_{-0.9}$}
\newcommand{\allzplain}{$10.5^{+1.0}_{-0.9}$}

\def\kms{\,km~s$^{-1}$}

\begin{document}

\title[Inhomogeneous reionization and globular clusters]{Evidence for inhomogeneous reionization in the local Universe from metal-poor globular cluster systems}
\author[L. Spitler et al.]
{Lee R. Spitler$^{1}$\thanks{E-mail: lspitler@astro.swin.edu.au} Aaron J. Romanowsky$^{2}$, J\"urg Diemand$^{3}$, Jay Strader$^{4,5}$,
\newauthor Duncan A. Forbes$^{1}$, Ben Moore$^{3}$ and Jean P. Brodie$^{2}$\\
$^{1}$Centre for Astrophysics \& Supercomputing, Swinburne University, Hawthorn, VIC 3122, Australia\\
$^{2}$University of California Observatories, Santa Cruz, CA 95064, USA\\
$^{3}$Institute for Theoretical Physics, University of Z\"urich, CH-8057 Z\"urich, Switzerland\\
$^{4}$Harvard-Smithsonian Center for Astrophysics, Cambridge, MA 02138, USA\\
$^{5}$Menzel Fellow
}

\date{Accepted 2012 March 31.  Received 2012 March 8; in original form 2011 September 8}

\pagerange{\pageref{firstpage}--\pageref{lastpage}} \pubyear{2012}
\maketitle

\label{firstpage}

\begin{abstract}
Exploiting a fundamental characteristic of galaxy assembly in the $\Lambda$CDM paradigm, the observed spatial biasing and kinematics of metal-poor globular star clusters are used to constrain the {\it local} reionization epoch around individual galaxies.  Selecting three galaxies located in different environments, the first attempt at constraining the environmental propagation of reionization in the local Universe is carried out.  The joint constraint from the three galaxies (\allz) agrees remarkably well with the latest WMAP constraint on $z_{reion}$ for a simple instantaneous reionization model.  More importantly, the range of $z_{reion}$ values found here are consistent with the global range of $z_{reion}$ estimates from other observations.  We furthermore find a $1.7\sigma$ indication that reionization completed in low-density environments before the intergalactic medium in high-density environments was reionized. This is consistent with certain theoretical models that predict that reionization was globally prolonged in duration, with neutral hydrogen pockets surviving in high-density environments, even after the surrounding regions were reionized.  More generally, this work provides a useful constraint on the formation history of galaxy stellar halos.
\end{abstract}


\section{Introduction}\label{intro}

Reionization marks the phase change of the Universe's intergalactic hydrogen gas from a neutral to an ionized state.  Though the general reionization picture follows from the fundamentals of Big Bang cosmology \citep{barkana_beginning:_2001}, a detailed timeline of how it propagated throughout the Universe is difficult to constrain because most of the direct observational signatures of this event are $\ga12$ billion light years distant from Earth.

Despite this challenge, observations show that the intergalactic medium was reionized no later than a redshift of $z\sim5$, not before $z\sim12$ and completed on a timescale of at least $\Delta z > 0.06$ \citep{fan_constraining_2006,ouchi_statistics_2010,robertson_early_2010,bolton_first_2010,bowman_lower_2010,larson_seven-year_2011}.  Though the dominant ionizing source contributing to reionization is still an open question, there is general agreement that the ionizing sources first appeared in high-density environments \citep{dijkstra_limit_2004,fan_constraining_2006,lidz_quasar_2007,wise_how_2008,power_primordial_2009,baek_reionization_2010,bunker_contribution_2010,robertson_early_2010,srbinovsky_fraction_2010,willott_canada-france_2010,yan_galaxy_2010,bouwens_lower-luminosity_2011,dopita_re-ionizing_2011,lorenzoni_star-forming_2011,mitra_reionization_2011,mitra_joint_2012}.

A key piece of information that is not yet constrained is the environmental propagation of reionization:  did reionization complete in dense environments first or were low-density voids the first locations to ionize?  While high-density environments of the Universe probably had a ``head-start'' and thus contained more ionizing sources, the same locations were also more likely to be shielded by dense clouds of dust and neutral hydrogen (HI) gas.  Furthermore, if the sources were predominately small galaxies, the UV background would have been more uniformly distributed than if the ionizing sources were typically active galaxy nuclei of rare, massive proto-galaxies.  Also, unlike ultraviolet radiation, X-rays produced by gas accretion onto black holes can escape local gas and dust absorption and ``pre-heat'' gas in lower-density regions \citep[e.g. ][]{baek_reionization_2010}.  Indeed, large-scale radiative transfer models show that the ionizing front can travel vast distances, from massive galaxies in dense regions to low-density locales \citep[e.g. ][]{weinmann_dependence_2007,finlator_late_2009,iliev_reionization_2011}. Understanding how reionization propagated through various environments will help constrain the properties of the ionizing sources and the state of the intergalactic medium during the reionization epoch.\\

One possible surviving relic of the reionization epoch may be the bimodal distribution of globular star clusters within large galaxies \citep{forbes_origin_1997,van_den_bergh_how_2001,beasley_formation_2002,santos_high-redshift_2003,rhode_metal-poor_2005,rhode_global_2007,rhode_wiyn_2010,moore_globular_2006,bekki_galactic_2005,bekki_origin_2008,griffen_globular_2010}:  systems of globular clusters show two distinct metallicity subpopulations \citep{zinn_globular_1985,gebhardt_globular_1999,neilsen_color_1999,larsen_properties_2001,kundu_new_2001,peng_acs_2006,strader_globular_2006,strader_globular_2007,kundu_bimodal_2007,spitler_extendingbaseline:_2008}.  The subpopulations include the ``metal-poor'' globular clusters (MPGCs) with typical metallicities a few percent of Solar metallicity and the ``metal-rich'' globular clusters (MRGCs) with metallicities $\sim30\%$ Solar.

Various theoretical works have invoked reionization as a way to produce the bimodal metallicity distribution in GC systems \citep[e.g. ][]{beasley_formation_2002,santos_high-redshift_2003,bekki_galactic_2005,moore_globular_2006,bekki_origin_2008,griffen_globular_2010}.  More specifically, these studies explored models where MPGCs form within small ($M_{\rm vir}\sim10^{8}M_{\odot}$) dark matter halos at $z\ga6$ \citep{bromm_formation_2002,boley_globular_2009}.  In this model, the MPGCs formed until cold molecular clouds within these halos was heated up during the reionization epoch.  This suppressed globular cluster formation for a period of time and allowed gases within the host galaxy to enrich in metals.  After some time had passed, a second generation of globular clusters, the metal-rich GCs, started to form during galaxy merger events \citep[e.g. ][]{ashman_formation_1992,beasley_formation_2002,griffen_globular_2010} and/or the growth of galaxy disks \citep{shapiro_star-forming_2010} at redshifts $z\sim2-4$ \citep{shapiro_star-forming_2010,spitler_first_2010}.  Although the global nature of the reionization epoch may provide a convenient way to explain the ubiquitous GC metallicity bimodality observed in nearby galaxies, a number of interesting alternative models have been proposed to explain GC metallicity distributions without reionization \citep[e.g. ][]{ct_formation_1998,scannapieco_triggering_2004,pipino_formation_2007,hasegawa_formation_2009,shapiro_star-forming_2010,muratov_modeling_2010,gray_formation_2010}.

\begin{figure}
\center
\includegraphics[scale=0.45]{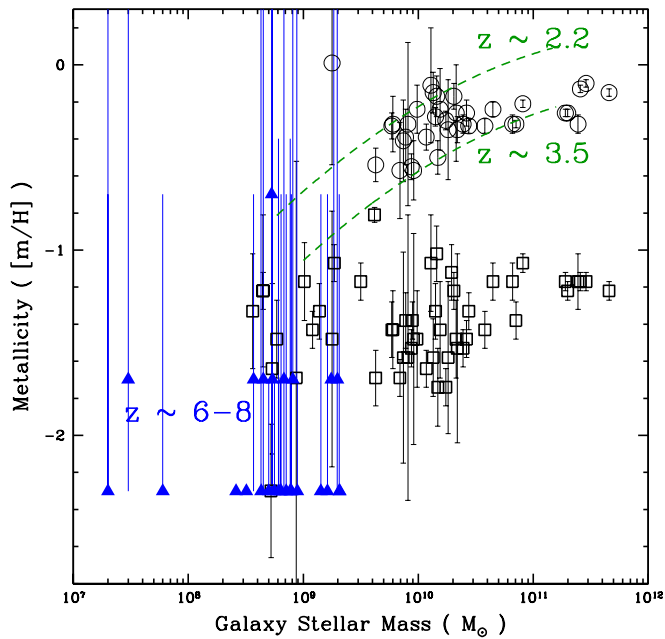}
\caption[Observations of metal-poor galaxies.]{Mean metallicity of metal-rich (circles) and metal-poor (squares) globular clusters in Virgo cluster galaxies as a function of their host galaxy stellar mass \citep[for details see ][GC data from \citealt{peng_acs_2006}]{spitler_first_2010}.  Metallicity measurements for galaxies at redshifts $z\sim6-8$ are given with measurement uncertainty ranges \citep[blue triangles; data from][see also the simulations of \citealt{salvaterra_simulating_2011}]{finkelstein_stellar_2010}.  Using metallicity as a timescale indicator, the overlap in stellar metallicity between metal-poor globular clusters (MPGCs) and the high-redshift galaxies supports the idea that MPGCs formed at redshifts $z\sim7-10$, when the \citet{finkelstein_stellar_2010} galaxies formed their stars. Dashed lines are observed mass-metallicity (from ionized gas observations) relations for redshifts $z\sim2.2$ and $z\sim3.5$ \citep{erb_stellar_2006,mannucci_lsd:_2009}.  As discussed in \citet{spitler_first_2010}, the overlap between these relations and the metal-rich globular clusters suggests they formed later than the MPGCs, at $z\sim2-4$.}\label{figmpgc}
\endcenter
\end{figure}

If reionization truncated globular cluster formation, the typical ages of metal-poor globular clusters should correlate with the {\it local} reionization epoch.  This means metal-poor globular clusters can be used to measure $z_{reion}$ for individual galaxies.  Current MPGC age estimates overlap with the expected reionization epoch.  Galactic MPGCs are constrained to be older than $\sim11$ Gyrs or $z\sim2$ \citep{krauss_age_2003}, with more stringent constraints on individual MPGCs suggestive of a pre-reionization epoch formation \citep[e.g. ][though exceptions do exist: \citealt{hansen_white_2007}]{hansen_hubble_2004}.  Constraints on the ages of individual extragalactic MPGCs are poorer \citep[$\ga 10$ Gyrs, see refs. in ][]{brodie_extragalactic_2006}, but are consistent with a pre-reionization formation.  Also, within the 1-sigma uncertainties the mean MPGC metallicities of various early-type galaxies are statistically consistent with preliminary measurements for galaxies at redshifts $z \sim 6-8$ \citep{finkelstein_stellar_2010,bouwens_very_2010,labbe_star_2010}.  This is shown in Fig.~\ref{figmpgc} and is consistent with a scenario where MPGCs were in place at redshifts greater than $z \sim 7$, well into the expected reionization epoch.

Unfortunately, the limited observational information about extragalactic globular clusters and large modeling uncertainties mean that direct methods to age-date extragalactic globular clusters do not provide a useful constraint on the local reionization epoch around individual galaxies.

A novel way to determine when the metal-poor GC formation epoch finished was recently proposed by \citet{diemand_distribution_2005} and \citet{moore_globular_2006}. The technique is based upon a fundamental property of a $\Lambda$CDM universe, where the most massive or rarest dark matter halos (and anything that formed within them) will tend to be centrally concentrated within their final host halos and show unique kinematic signatures at the present epoch.  Thus by constraining the spatial and kinematic properties of metal-poor globular clusters today, information about their progenitor halos can be recovered.  Since the rarity of a halo for a given virial mass depends on redshift, this information can be used to age-date metal-poor GCs with enough accuracy to constrain the reionization epoch around the galaxy hosting the GCs \citep[][see also \citealt{bekki_galactic_2005}]{diemand_distribution_2005,moore_globular_2006}.

This technique was demonstrated in \citet{moore_globular_2006} using the Milky Way's globular cluster system.  They found evidence for a reionization redshift of $z_{reion}=10\pm2$ \citep[updated for a WMAP7 cosmology,][]{komatsu_seven-year_2011}.  In the present work, this technique is applied to two additional galaxies located in denser environments than the Milky Way to help constrain the temporal propagation of reionization through the local Universe.\\

Section~\ref{theoryintro} provides a detailed background, describing in detail both the theoretical framework of the ``Diemand-Moore'' methodology and relevant globular cluster observations. In Section~\ref{observations}, the observational data are presented. The main mass modeling and $z_{reion}$ measurements are presented in Section~\ref{method}.  It is followed by a discussion in Section~\ref{discussion}.

\section{Method and Assumptions}\label{theoryintro}

It has been shown that the growth of structure in the Universe can be thought of as a Gaussian mass-density field where mass-density fluctuations form and assemble hierarchically \citep{press_formation_1974,lacey_merger_1993}.  The rare mass-density fluctuations in the Universe are biased towards high-density regions of the Universe \citep{cole_biased_1989,sheth_large-scale_1999}.  As a consequence, the rarest fluctuations will tend to be the most centrally-concentrated material in a galaxy \citep{moore_resolving_1998,white_where_2000}.  By convention, the height of the fluctuations are represented by $\nu$, the number of standard deviations above the mean mass-density level at that epoch.

\citet{diemand_distribution_2005} used N-body simulations to explicitly show that the location and kinematic properties of mass structures within a galaxy today depend on the rarity of the mass structure or ``halo'' when it reached its Jeans mass and gravitationally collapsed.  They found that the rarest halos are more likely to accrete onto a galaxy at earlier times, when the gravitational potential of the galaxy was relatively small.  Thus any object that formed within a rare halo is more likely to have a smaller infall velocity and will occupy the central regions of the galaxy until the present epoch.  This means that an object's location and orbit or kinematics within a halo today can be used to recover the properties of the halo it formed within \citep{diemand_distribution_2005,moore_globular_2006}.  Examples of ``tracer'' objects include a galaxy's satellites, its stellar halo, remnants of population III stars, and metal-poor globular clusters.\\

The ``Diemand-Moore'' method is a way use the spatial concentration of MPGCs relative to the hosting galaxy's halo mass profile to constrain the rarity, $\nu\sigma$, of the halos that the MPGCs formed within.  Results from computer simulations are used to translate the spatial bias of the MPGCs into a $\nu\sigma$ constraint.  The $\nu\sigma$ value alone does not constrain the redshift when MPGCs stopped forming, so an additional assumption is made to break the degeneracy between redshift and the typical halo mass for a given $\nu$.  By invoking the argument that star formation only occurs efficiently in halos with virial temperature more than T$_{\rm vir}=10^4$ K, the degeneracy can be broken and the redshift when MPGCs were suppressed by reionization is found.  More details are provided in the following Sections.

The Diemand-Moore method requires at least three ingredients to constrain the $\nu\sigma$ value of MPGC progenitor halos:  MPGC surface density profile from observations, a halo mass model for the galaxy hosting the MPGCs and a theoretical framework \citep[here ][is used]{diemand_distribution_2005} to interpret the spatial bias of the MPGCs, relative to the host galaxy's mass profile.  MPGC kinematic information from line-of-sight velocity measurements are an optional 4th ingredient that can be used for an additional, albeit weaker, constraint on $\nu\sigma$. The kinematics also prove to be a useful tool to constrain the galaxy mass model in a self-consistent manner to the theoretical framework.

The main physical model that the following analysis depends upon is outlined here:
\begin{itemize}
\item Each MPGC formed at high redshift in a small, rare dark matter halo (likely together with other MPGCs and a small proto-galaxy). 

\item The bath of UV radiation associated with the reionization epoch was such that MPGC formation was ``instantaneously'' suppressed\footnote{If a locally-instantaneous reionization model is incorrect and an extended reionization is preferred, the MPGC constraints still provide a lower-limit on $z_{reion}$.  This is because the MPGCs will tend to be associated with the most abundant class of halo at the end of the extended reionization epoch:  the least-massive or lowest $\nu\sigma$ halo that has only just exceeded T$_{\rm vir}=10^4$ K and can form MPGCs.  The final spatial distribution of MPGCs will therefore reflect this low $\nu\sigma$ class of halos and hence the epoch when reionization completed can still be recovered.} in the GC formation sites.  This allows one make a simple link between the MPGCs and a single $\nu\sigma$ class of halos, thereby constraining $z_{reion}$ exactly.

\item The halo containing MPGCs was later accreted by the host galaxy halo, i.e. it became a subhalo. 

\item  This accretion process effectively disrupts the early accreted, biased, rare dark matter subhalos, but the globular cluster tracers remain intact and will maintain the spatial and kinematic properties of the subhalos even to the present day \citep{diemand_distribution_2005}.

\end{itemize}

Should the second assumption prove to be incorrect, the analysis still provides an interesting characterisation of the progenitor subhalos of MPGCs and perhaps the other mechanism(s) that truncated their formation.  Also, the tracers need not be MPGCs and can include e.g. the Milky Way stellar halo \citep{diemand_distribution_2005} and its satellite galaxies \citep[][see also \citealt{ocvirk_signature_2011}]{moore_globular_2006}.  Indeed, a useful feature of the Diemand-Moore technique is that it does not care about the detailed baryon physics of e.g. star formation, but only requires an observable tracer of subhalos that accreted relatively early in a galaxy's formation history.

\section{Target Galaxies and the Observations}\label{observations}

In this Section, the observational data are described.  To constrain $\nu\sigma$ for MPGC progenitors in a galaxy, the MPGC surface density profile must be measured out to large radii (typically tens of arcminutes), where contamination starts to contribute significantly to the profile.  To reduce contamination levels, high-quality imaging is needed.  To statistically subtract the remaining contamination, spectroscopic information can be used \citep[e.g. ][]{strader_wide-field_2011} or the profile must extend to large enough radii to characterise contamination levels.  Furthermore, the galaxy must have relatively ``well-behaved'' kinematics so that an estimate of the mass model can be derived.

The extragalactic targets described here are the only galaxies that currently meet these demanding requirements.  The galaxies also conveniently reside within different galaxy environments, including the ``high-density'' environment of a cluster galaxy, a ``medium-density'' group and a ``low-density'' field galaxy.  

\subsection{Messier 87}\label{m87}

Messier 87 (M87) is the central galaxy associated with the Virgo cluster of galaxies. It is located 16.5 Mpc away and the cluster contains hundreds of member galaxies. Though M87 and the Virgo cluster are frequently treated as the same thing (at least in terms of their ``common'' dark matter halo), recent analysis supports early suggestions that the Virgo cluster is not a single massive galaxy cluster halo, but is instead made up of a set of subhalos in an unrelaxed state \citep{binggeli_studies_1987,strader_wide-field_2011}.  The M87 galaxy-sized halo is thought to be virialized, but is a distinct subhalo from the rest of the Virgo structure \citep[see discussion in ][]{doherty_edge_2009,strader_wide-field_2011,romanowsky_ongoing_2011}.

From kinematic tracers within $\sim100$ kpc, the extrapolated virial mass (\mh) of the M87 halo is $M_{\rm vir}\sim10^{14}$~\msun~\citep{strader_wide-field_2011}.  The surrounding Virgo cluster halo is \mh~$\sim10^{15}$~\msun \citep[e.g. ][]{mclaughlin_evidence_1999,strader_wide-field_2011}, which means M87 resides in high-density environment and was therefore one of the first locations in the local Universe to host star formation.  It has even been proposed that ionizing radiation from sources in Virgo had enough of a head-start to travel to the Local Group and reionize the Milky Way galaxy \citep[e.g. ][]{weinmann_dependence_2007}. For these reasons, M87 and the Virgo cluster are a particularly interesting location to measure the local reionization epoch.

The M87 GC catalogue is presented in \citet{strader_wide-field_2011} and was constructed from archival CFHT/Megacam \citep{boulade_megacam:_2003} images.  Kinematic information for the GCs come from \citet{strader_wide-field_2011}, who present new Keck/DEIMOS \citep{faber_deimos_2003}, Keck/LRIS \citep{oke_keck_1995} and MMT/Hectospec \citep{fabricant_hectospec_2005} line-of-sight velocities.  Out of an estimated MPGC population of $\sim14000$ MPGCs \citep{strader_wide-field_2011}, 289 have useful velocity information.  

The M87 MPGC surface density was derived in \citet{strader_wide-field_2011} using standard reduction techniques and methodologies.  Briefly, GCs were photometrically selected down to $i=22.5$ mag. over a region extending to radii $\sim130$ kpc from M87.  Hubble Space Telescope imaging and the velocity information were used to accurately constrain the amount of contamination present in the GC catalogue, which was subtracted off the MPGC surface density profile before analysis.

\subsection{NGC~1407}\label{n1407}

The second target of this study is an elliptical galaxy named NGC~1407, whose galaxy group contains $\sim30$ members \citep{brough_eridanus_2006}.  It is 21 Mpc distant and has a sizable GC system \citep{forbes_imaging_2006,harris_globular_2006}.  \citet{romanowsky_mappingdark_2009} found the galaxy group, which is centred on NGC~1407, to have a relatively high mass-to-light ratio for its total virial mass \mh $\sim6\times10^{13}$ \msun.  This target provides a probe of reionization in a ``medium-density'' environment.

Two Subaru/Suprime-Cam \citep{miyazaki_subaru_2002} image datasets of NGC~1407 were analysed.  Each covers a $\sim34\arcmin\times34\arcmin$ area on the sky.  The first is centered on NGC~1407 and is made up nearly 5 hours of $g,r,i$ imaging under $0.5-0.6\arcsec$ conditions.  The resulting GC catalogue has been used for an extensive spectroscopic campaign with Keck/DEIMOS \citep{romanowsky_mappingdark_2009,foster_deriving_2010,pota_survey_2011} and a photometric study \citep{forbes_evidence_2011}.  The second Suprime-Cam dataset is an archival dataset in $B,V,I$ bands with $0.6-0.7\arcsec$ seeing and totaling $\sim2$ hours in exposure time.  In this imageset,  NGC~1407 was positioned toward one end of the mosaic so the radial coverage extends $\sim30\arcmin$ from NGC~1407 or $\sim200$ kpc.  The GC catalogues are available upon request from the first author.

MPGC surface density profiles were constructed in each dataset from a catalogue of GC candidates found in the images (MPGCs were taken to have $(g-i)_0\le0.98$).  A literature Hubble Space Telescope, Advanced Camera for Surveys catalogue \citep{forbes_imaging_2006} was incorporated into the analysis to improve the surface density profile in the central $\sim3\arcmin$ of the galaxy center.  Standard procedures were followed to derive the GC surface density profile \citep[e.g. ][]{spitler_connection_2008}.  The surface density data extend to $\sim210$ kpc and a constant surface density is not reached at large radii, suggesting a small number of MPGCs should be found beyond the region covered by the Suprime-Cam imaging.  Since contamination was not subtracted from the MPGC surface density profile, the fits to the NGC~1407 MPGCs described in subsequent sections include a constant background level to model the contamination.

An expanded GC line-of-sight velocity catalogue from \citet{romanowsky_mappingdark_2009} and \citet{foster_deriving_2010} is used, which incorporates new Keck/DEIMOS observations \citep{pota_survey_2011}.  The total number of NGC~1407 MPGCs with spectroscopic velocities is 167 GCs (excluding suspected ultra compact dwarf candidates; see details below), out of an estimated total of $\sim4000$.

\subsection{Milky Way}\label{mw}

The Milky Way (MW) MPGC spatial and kinematic information come from the new \citet{harris_new_2010} catalogue.  The MPGC sample contains 110 GCs with $[Fe/H]<-0.9$ and has a mean metallicity of $[Fe/H]\sim-1.6$.  The Milky Way GC analysis is carried out in three-dimensions rather than in projection as for the preceding galaxies.  In order to model the halo mass profile of the Milky Way with the MPGC velocity information, heliocentric distances and velocities must be corrected to the Galactic reference frame.  The distance correction is straightforward and included in the \citet{harris_new_2010} catalog itself, but the velocities in principle require proper motion measurements, and these are in many cases not available with sufficient precision.  Instead, after correcting the line-of-sight velocities for the heliocentric motion (e.g., eq. 5 of \citealt{xue_milky_2008}), the data are corrected to the true radial velocities on a statistical basis:  given the anisotropy profile $\beta(r)$ from the best-fit model (discussed in the next Section), the correction from line-of-sight to radial velocity dispersion (eq. 1 of \citealt{battaglia_erratum:_2006}) is applied to the observed velocities, and the dispersion profile is calculated.  This processes is repeated iteratively and ultimately yields a very small correction that affects the final $\nu$ values at only the $\sim$~3\% level (after excluding from the analysis the GC kinematics inside $r=$~8~kpc, which are more difficult to correct).

\section{Analysis}\label{method}

The Diemand-Moore technique uses the spatial bias of MPGCs relative to the hosting galaxy's dark matter distribution to constrain the $\nu\sigma$ properties of the MPGC progenitor subhalos.  This is accomplished by fitting a modified \citet[][ NFW]{navarro_structure_1996} function from \citet{diemand_distribution_2005} to the MPGC spatial distribution.  The fit also requires as input the scale radius ($r_s$) of the host galaxy halo, and thus an accurate mass model of each galaxy is an important component of this work.  Although models for these galaxies exist in the literature, they are derived using different techniques and various mass tracers.  To remove possible systematic errors resulting from these inhomogeneities, new mass models are derived for each galaxy.  Also, to ensure the $\nu\sigma$ fits to the surface density profile are derived in a self-consistent manner with the galaxy mass models, MPGC observations are used to derive the mass model.

The mass modelling methodology is outlined in the first subsection below.  In subsequent subsections, the individual galaxy mass models are derived.  Though some necessary assumptions and compromises are made to yield the final mass models, the halo mass properties and their uncertainties are robust.  In the final subsection, \S\ref{reion}, the $\nu\sigma$ measurements for each are converted into $z_{reion}$ constraints. 

In the following, the MPGC spatial profile is denoted $j(r)$ and $\Sigma(R)$, for three-dimensional and the projected cases, respectively.  Also, when $\nu\sigma$ constraints from the Diemand-Moore technique are discussed, $\nu$ actually refers to all subhalos {\it above} the subhalo virial mass corresponding to that $\nu$.  For example, the constraint \mwnuss~for the Milky Way, actually includes all subhalos equal to and more massive than a \mwnuss~halo at the redshift of interest.

\subsection{Derivation of the Galaxy Mass Models}\label{massmodels}

\begin{figure*}
\includegraphics[scale=0.7]{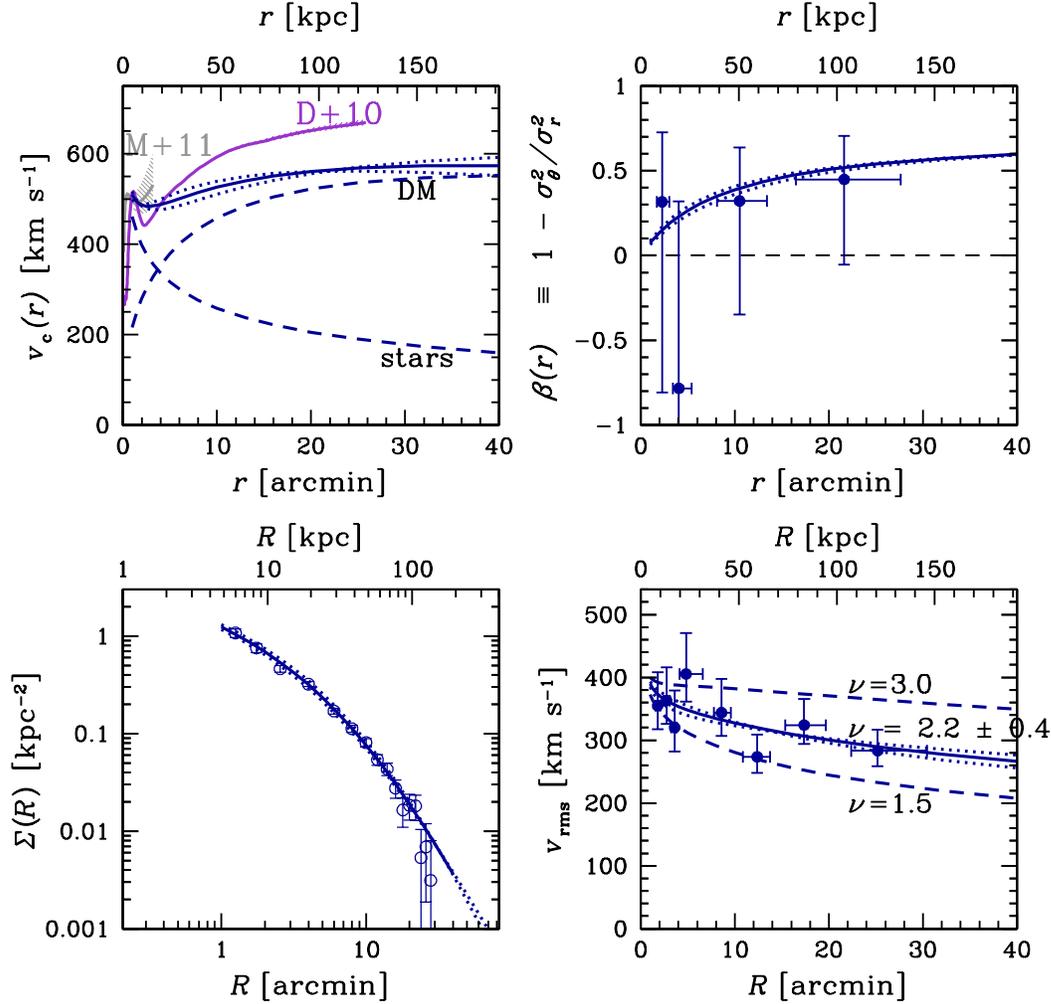}
\caption[m87disp]{
Modeling ingredients 
for the M87 metal-poor globular cluster system.
Model predictions are shown as curves:
the solid curves are the best model, with 
$\nu=2.19$, $r_s=$~126~kpc, and $\rho_s= 1.4\times10^6 M_\odot$~kpc$^{-3}$
(a normal concentration halo with $M_{\rm vir}\sim4.8\times10^{13}M_\odot$).
The dotted curves are
$\nu=2.61$, $r_s=$~191~kpc, and $\rho_s=0.7\times10^6 M_\odot$~kpc$^{-3}$
(a low concentration halo with $M_{\rm vir}\sim7.2\times10^{13}M_\odot$);
and 
$\nu=1.80$, $r_s=$~85~kpc, and $\rho_s= 2.9\times10^6 M_\odot$~kpc$^{-3}$
(a high concentration halo with $M_{\rm vir}\sim3.4\times10^{13}M_\odot$).
Note, the mass modelling only incorporates tracers from within $\sim100$ kpc of M87 so the halo properties are a reflection of the M87 subhalo, not the entire Virgo cluster halo \citep[e.g. see][]{doherty_edge_2009,strader_wide-field_2011}. 
$R$ is for projected radii, $r$ is for physical 3D radii.
{\it Upper left:} Circular velocity profile.
The total profile for the best model is shown decomposed into its stellar and
dark matter components (dashed curves, as labeled).
Independent results are also shown from stellar dynamics \citep{murphy_galaxy_2011}
and X-ray gas \citep{das_steepening_2010}.
{\it Upper right:} MPGC velocity anisotropy in the model (lines) and estimated from the kurtosis of the observed line-of-sight velocity distribution (circles).
{\it Lower left:} MPGC surface density.
{\it Lower right:} MPGC line-of-sight velocity dispersion.
Data are shown by points with error bars \citep{strader_wide-field_2011}.
In fitting the dispersion data, there is a strong degeneracy between the halo parameters
$r_s$ and $\rho_s$, which is broken only by imposing cosmological priors.
The remaining uncertainty on $r_s$ translates directly into an uncertainty on $\nu$.
The dashed curves show two models that are ruled out by the dispersion data,
even after allowing for a 1~$\sigma$ scatter in the concentration.
}\label{fig:m87disp}
\end{figure*}

Here we outline our modeling methods that will be used for specific galaxies
in the following subsections.
We begin with the three-dimensional number density profile of the
dynamical tracer population (MPGCs in this paper).
This profile is modeled as a modified NFW profile, as derived in
\citet{diemand_distribution_2005}:
\begin{equation}\label{equ_diemand}
\rho(r,\nu) = \frac{\rho_s}{(r/r_{\nu})^{\gamma}[1 + (r/r_{\nu})^{\alpha}]^{(\beta_{\nu}-
\gamma)/\alpha} },
\end{equation}
where $\nu$ is the rarity value, $\gamma$ and $\beta_{\nu}$ are the inner and outer
power-law slopes, respectively, and $\alpha$ describes the intermediate behavior.
Following \citet{diemand_distribution_2005}, we set $\alpha=1$,
$\gamma=1.2$, and $\beta_{\nu} = 3 + 0.26\nu^{1.6}$.  Thus a higher $\nu$ value implies a steeper outer slope.  
The value of $\nu$ is related to a characteristic radius $r_\nu$ by:
\begin{equation}\label{eqn:nu}
\nu \equiv 2 \ln\left(r_s/r_\nu\right) ,
\end{equation}
which means the surface density is more compact for higher values of $\nu$.  

Ideally, we would compare this model to an observed surface density profile,
and fit for $\nu$.  However, in practice there are not good prior constraints on
$r_s$ available, i.e., we do not know the DM distribution of the halo which we
need to estimate the relative compactness of the MPGC subpopulation.\footnote{One
might think that with massive ellipticals like NGC~1407 and M87, 
an X-ray based mass profile would provided the much-needed independent information
on the DM profile.  However, as discussed in \cite{romanowsky_mappingdark_2009},
for these particular galaxies,
there are inconsistencies between the X-ray results and GC system dynamics.
In fact, there is an emerging, wider pattern of disagreement between X-ray and optical mass determinations in elliptical galaxies (e.g., \citealt{shen_supermassive_2010}).
For this paper, it is assumed that the MPGC dynamics results are reliable.}
To solve this problem, we will make use of the fact that there are {\it kinematical}
predictions of the model, so we can use the observed spatial and kinematical
distributions to fit simultaneously for $\nu$ and $r_s$.

Our basic approach for the kinematics is to take model distributions for
mass, MPGC density, and anisotropy, solve a spherical Jeans equation (e.g., \citealt{mamon_dark_2005}) and derive a velocity dispersion profile for the MPGCs, $\sigma(r)$.  
In the Milky Way, the $\sigma(r)$ prediction can be directly compared to 
observational data, while in external galaxies, an additional step is needed
to project to line-of-sight dispersions, $\sigma_{\rm p}(R)$.

Each mass model consists of a stellar component (with a reasonable value for
the stellar mass-to-light ratio $\Upsilon_*$) and a $\Lambda$CDM-based model for the DM
distribution. This DM profile has a similar parametric form to Equation~\ref{equ_diemand},
but with density and scale parameters ($\rho_s$ and $r_s$) that are highly correlated.
To derive these correlations (or modeling priors), we 
begin with the relation between virial mass and concentration predicted for relaxed halos at $z=0$ by the Bolshoi and BigBolshoi simulations using WMAP5 cosmological parameters \citep[eq. 25]{prada_halo_2011}.  After converting the mass and concentration into scale radius and density, and subtracting off a cosmological baryon fraction of $0.17$ to find the DM density alone, the following power-law provides an approximation to the average results of the numerical simulations:
\begin{equation}\label{eqn:rhosrs}
\rho_s = 1.16\times10^7 \, \frac{M_\odot}{{\rm kpc}^3} \left(\frac{r_s}{\rm kpc}\right)^{-0.42} ,
\end{equation}
which is good to 8\% or better over the virial mass range of $\sim 10^{11}$--$10^{15} M_\odot$.  A 1~$\sigma$ cosmological scatter in halo concentration of $\pm0.11$~dex \citep{duffy_dark_2008,maccio_concentration_2008} translates to a scatter in $\rho_s$ at fixed $r_s$ of $\pm0.24$~dex.  This scatter ultimately provides the largest systematic uncertainty on the measurements of $\nu$ and hence $z_{reion}$.

These results involve fits of standard NFW density profiles to the simulations
(i.e., Equation~\ref{equ_diemand} with $\gamma=1$, $\alpha=1$, $\beta=3$).  
However, the \citet{diemand_distribution_2005} analyses used a slightly different DM density model, where the central cusp has a power-law slope of $\gamma=-1.2$ rather than $-1$:
\begin{equation}
\rho(r) = \rho_s \left(\frac{r}{r_s}\right)^{-1.2} \left(1+\frac{r}{r_s}\right)^{-1.8} .
\end{equation}
Here it is assumed if such a model had been fitted to the recent simulations, the $M_{\rm vir}$ and $r_s$ values would be the same as with the NFW fits.  Eq.~\ref{eqn:rhosrs} then becomes:
\begin{equation}\label{eqn:rhosrs2}
\rho_s = 1.07\times10^7 \, \frac{M_\odot}{{\rm kpc}^3} \left(\frac{r_s}{\rm kpc}\right)^{-0.42} .
\end{equation}
In practice, the increased fractional mass inside $r_s$ means that lower $\rho_s$ values will be fitted to the data, relative to an NFW model.  The model for a mass-sequence of average DM circular velocity profiles is then:
\begin{multline}\label{eqn:vc}
v_{\rm c,DM}(r) = 17.9 \, \frac{\rm km}{\rm s} \left(\frac{r_s}{\rm kpc}\right)^{0.79} \left(\frac{r}{r_s}\right)^{0.4} \\
\left[{}_2F_1\left(1.8,1.8;2.8;-\frac{r}{r_s}\right)\right]^{1/2} ,
\end{multline}
where ${}_2F_1$ is a hypergeometric function.

The final ingredient in the models is the velocity dispersion anisotropy profile,
which is normally a major source of uncertainty in modeling observations of
pressure-supported galaxies, but in the context of the Diemand-Moore models,
is uniquely constrained.  This anisotropy profile is:
\begin{equation}\label{eqn:betar}
\beta(r) = \beta_0 \frac{r}{r+r_\nu} ,
\end{equation}
where:
\begin{equation}
\beta_0 = 1 - \frac{2}{5\sqrt{\nu}} .
\end{equation}
This anisotropy profile is isotropic in the center ($\beta\simeq0$ for $r \ll r_\nu$) and becomes radially biased in the outer regions ($\beta \simeq \beta_0$ for $r \gg r_\nu$).  Higher values of $\nu$ imply stronger radial anisotropy.

A complication for modeling both NGC~1407 and M87 is that there are some correlations between GC luminosities and kinematics that are probably driven by populations of ultra-compact dwarfs (UCDs) at the bright end, with distinct dynamics from the ``normal'' GCs \citep[see ][for more details]{romanowsky_mappingdark_2009,strader_wide-field_2011,brodie_relationships_2011}.  Although the UCDs comprise only a tiny part of the overall ``GC'' systems, they represent a substantial fraction of the luminosity-based spectroscopic samples.  Such objects were therefore eliminated from the kinematic samples by excluding anything that has a large measured size (half-light radius $>5$~pc), or that has no size measurement but is bright ($i_0 < 21.3$ and $i_0 < 20$ for NGC~1407 and M87, respectively).  These objects have no impact on the measured surface density profiles.

The following subsections present the details of the modeling for each
galaxy, and the resulting $\nu$ constraints.

\subsection{Rarity of M87 MPGC progenitor subhalos}

For M87, the mass models are tuned not only to fit the MPGC subpopulation velocity dispersion but also to have a total $v_{\rm c}\simeq$~500~\kms\ at $r\sim$~5~kpc for consistency with the stellar dynamics results of \citet{murphy_galaxy_2011}; this means adjusting $\Upsilon_*$ accordingly.  As discussed in \citet{strader_wide-field_2011}, the NFW models have some difficulty in reproducing the stellar dynamics inferences at $r\sim$~15~kpc; the cuspier \citet{diemand_distribution_2005} models fare better, but there may still be some tension with the stellar data.  These mass models are also completely unable to reproduce the X-ray based mass-model of \citet{das_steepening_2010} while maintaining strongly radial anisotropy and reproducing the observed velocity dispersions (an additional example of the problem is footnoted in \S\ref{massmodels}).

Figure~\ref{fig:m87disp} shows the elements of the modeling procedure, along with comparisons to data as applicable.  The model consists of profiles of $v_{\rm c}(r)$, $\beta(r)$, and $j(r)$ (the latter is not shown), which are then input to the Jeans equation to solve for $\sigma_r(r)$ and $\sigma_\theta(r)$, and projected to $\Sigma(R)$ (lower left panel in Fig.~\ref{fig:m87disp}) and $v_{\rm rms}(R)$ (lower right panel).  As a consistency check, the model expectations for $\beta(r)$ are compared with an empirical estimate from the line-of-sight velocity distribution of the data in the upper right panel of Fig.~\ref{fig:m87disp}.  This estimate makes use of a restricted class of spherical constant-$\beta$ models which predict the line-of-sight velocity {\it kurtosis} if the velocity dispersion profile is approximately constant with radius \citep{napolitano_planetary_2009}.  The kurtosis measurements from the MPGC data can thus be converted to rough estimates of the anisotropy profile.

For M87, a typical-concentration $\Lambda$CDM halo with $r_s=126$~kpc ($M_{\rm vir} \sim 5\times10^{13} M_\odot$) is found, along with the \citet{diemand_distribution_2005} anisotropy prediction, to fit the M87 MPGC dispersion data remarkably well.  The implied $\beta(r)$ is also nicely consistent with the empirical estimates.  This model corresponds to $\nu=2.19$.  The predicted velocity dispersions are fairly sensitive to the value of $r_s$, increasing for larger $r_s$ (and $\nu$).  Using a $\chi^2$ fit to the dispersion data, $r_s$ is determined at the $\pm10$~kpc level, and $\nu$ at the $\pm0.08$ level.

These results hold for the strong theoretical $\Lambda$CDM prior adopted on the halo parameters.  Relaxing this prior allows for a much larger range of solutions, as various combinations of halo density and scale radius can all provide similar mass profiles in the region with MPGC observations.  Allowing for the predicted 1~$\sigma$ scatter in the halo properties (i.e., varying the normalization of Eq.~\ref{eqn:vc} by $\pm$~0.12~dex), $r_s$ is constrained at the $\sim$~50\% level, yielding \mesnu~(see Figure~\ref{fig:m87disp}).

\subsection{Rarity of NGC~1407 MPGC progenitor subhalos}

For NGC~1407, several priors are invoked: the mass-concentration relation from $\Lambda$CDM; an estimate of the mass at $R\sim$~100~arcmin from satellite galaxy dynamics ($v_{\rm c} \sim 700\pm 100$~\kms; see \citealt{romanowsky_mappingdark_2009}); and a measurement of the central stellar velocity dispersion around 1~$R_{\rm e}$ of $\sigma_{\rm p} \simeq 260$~\kms\ \citep{proctor_probing_2009}.  It is beyond the scope of this paper to carry out the detailed dynamical modeling required to convert this $\sigma_{\rm p}$ measurement into an estimate of $v_{\rm c}$, and instead the empirical results from dynamical modeling of various other pressure-supported galaxies that $v_{\rm c}/\sigma_{\rm p} \sim$~1.4--1.7 \citep{wolf_accurate_2010,trujillo-gomez_galaxies_2011,murphy_galaxy_2011,dutton_dark_2011} are used.

\begin{figure*}
\includegraphics[scale=0.7]{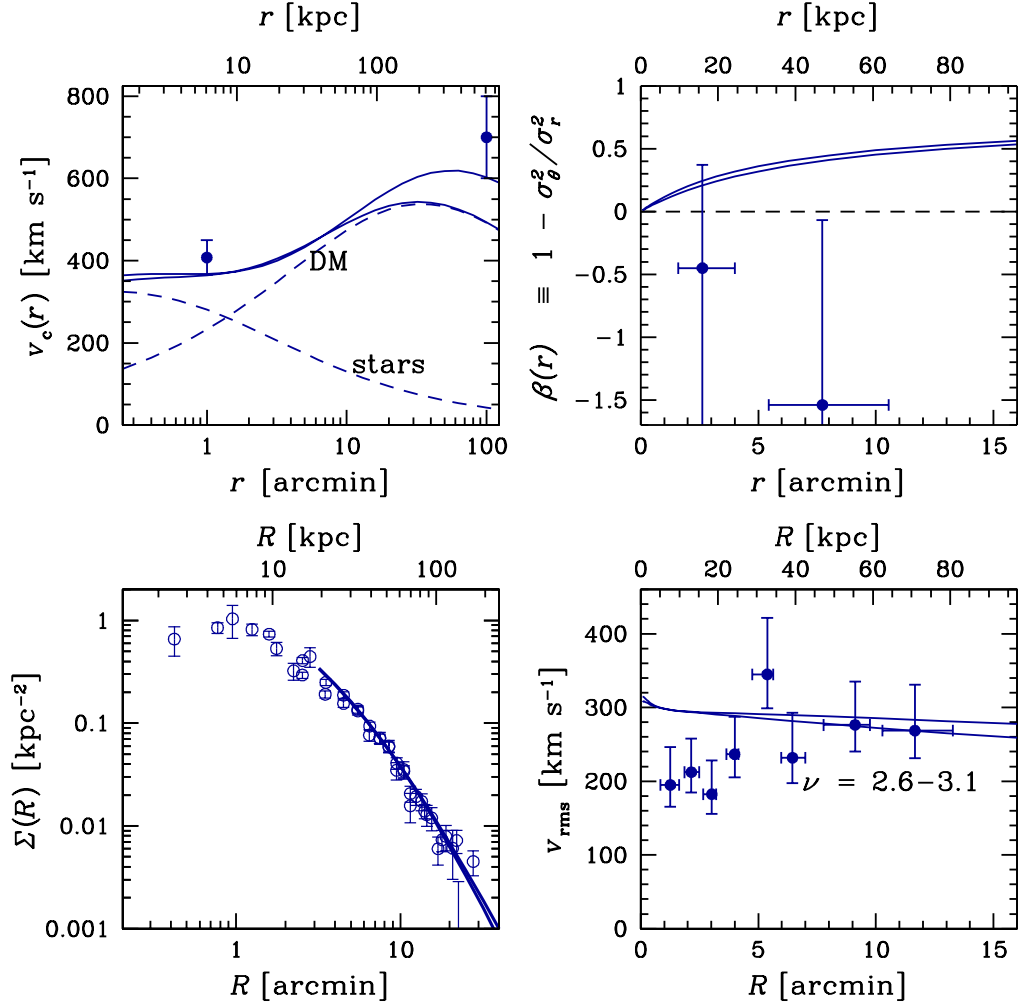}
\caption[n1407disp]{
As Figure~\ref{fig:m87disp}, for NGC~1407.
Error bars in the upper-left panel are mass profile constraints from stellar and satellite galaxy dynamics,
as discussed in the text.
Two possible solutions are shown by solid curves:
one is a normal-concentration halo with $\nu=2.61$,
$r_s=122$~kpc, $\rho_s=1.4\times10^6 M_\odot$~kpc$^{-3}$,
and $M_{\rm vir}\sim4.4\times10^{13}M_\odot$;
the other is a low-concentration halo with $\nu=3.11$,
$r_s=205$~kpc, $\rho_s=0.7\times10^6 M_\odot$~kpc$^{-3}$,
and $M_{\rm vir}\sim8.6\times10^{13}M_\odot$.
}\label{fig:n1407disp}
\end{figure*}

Unlike M87, for NGC~1407 there is no model solution that straightforwardly fits all of the above constraints (see Figure~\ref{fig:n1407disp}).  In particular, it appears to be very difficult to match the inner MPGC velocity dispersion profile, where the $\sigma_p \sim$~200~\kms\ inside $\sim$~4~arcmin ($\sim$~25~kpc) is lower than the stellar velocity dispersion.  Normally, the MPGC dispersion should be {\it higher} because the spatial distribution is more extended.  Lowering the dispersion would require {\it tangential} orbital anisotropy, which contradicts expectations in the \citet{diemand_distribution_2005} model for mild radial anisotropy.  It is speculated instead that GC disruption processes near the galactic center have preferentially depleted the more radial orbits \citep[e.g. ][]{fall_dynamical_2001,vesperini_modeling_2003}.  MPGCs inside 3.5~arcmin ($\sim20$~kpc) are therefore excluded from the analysis (it is not clear why this should be necessary for NGC~1407 but not for M87).

Considering now the outer MPGC velocity dispersion data, it also turns out to be difficult to fit the joint constraints from $\Lambda$CDM halo concentration expectations and the satellite galaxies dynamics.  The mass required by the latter predicts MPGC dispersions higher than observed, which can be alleviated by a low-concentration halo.  The range of plausible solutions is bracketed by adopting a halo that has a low concentration at the 1~$\sigma$ level while fitting the satellites constraint, or one with a normal concentration that does not agree with the satellites (whose constraint is then assumed to be unreliable).  Alternative modifications to lower the MPGC dispersions would be to use a less cuspy NFW model (which is within the cosmological scatter for halo profiles), or to decrease the adopted distance since the kinematics data will then effectively be probing less far into the halo (although this galaxy is likely to be {\it more} distant, not less, based on estimates summarized in NED).

The resulting range in fit parameters is $\nu\sim$~2.6--3.1; again, the systematic uncertainties outweigh the statistical ones ($\sim \pm$~0.15) from fitting the dispersion and density profiles.  An additional issue from these solutions is that the low observed kurtosis for the outer MPGCs suggests tangential anisotropy in these regions too, again at odds with the radial model expectations.

\subsection{Rarity of Milky Way MPGC progenitor subhalos}

\begin{figure*}
\includegraphics[scale=0.7]{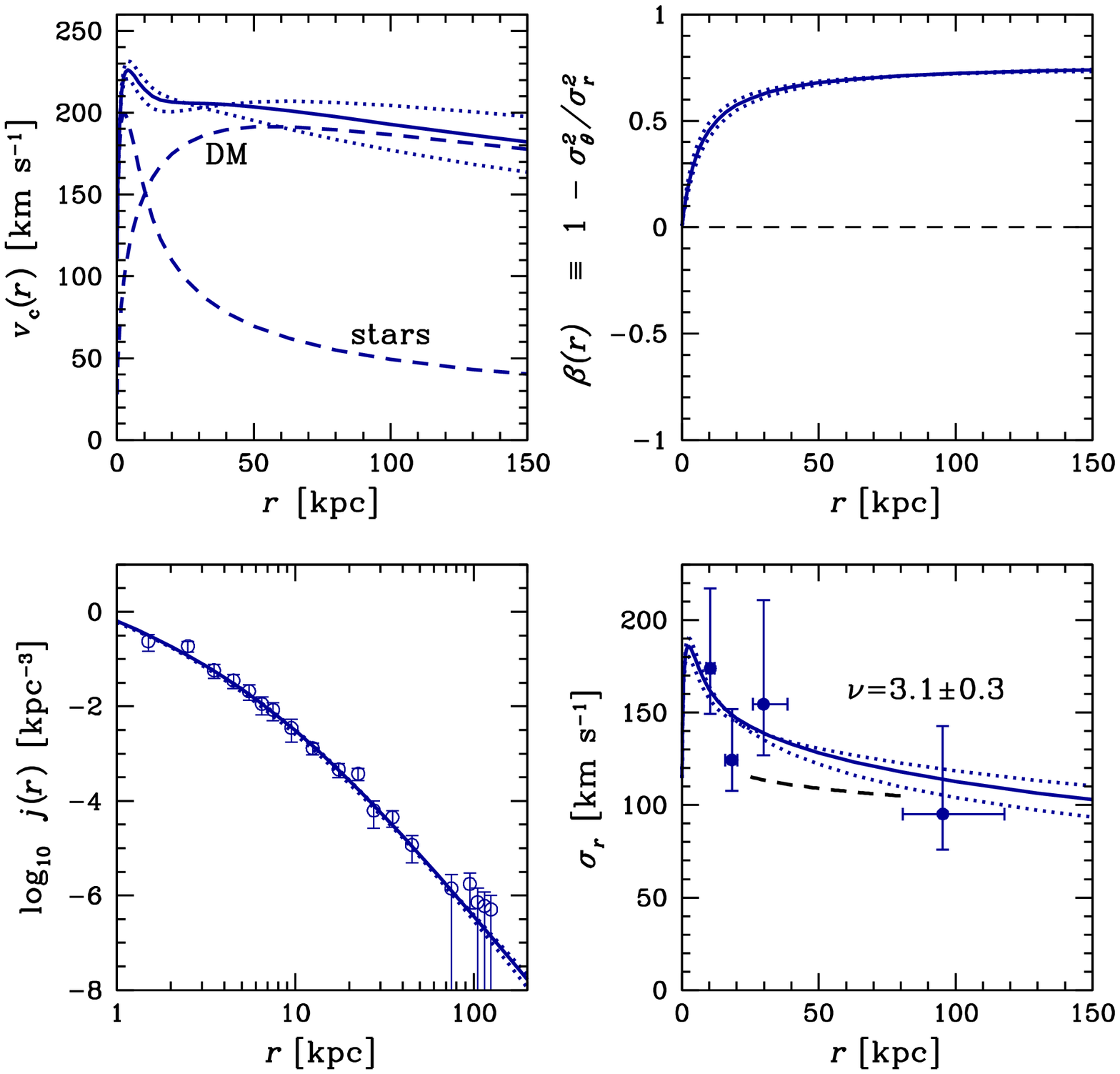}
\caption[mwdisp]{
As Figure~\ref{fig:m87disp}, for the Milky Way.
Here the lower panels are three-dimensional rather than projected quantities.
The dashed curve in the lower-right panel shows the velocity dispersion profile
fit to halo tracers (mostly stars) from \citet{gnedin_mass_2010}.
The solid curve in this panel shows the best model, with $\nu=3.12$,
$r_s=33$~kpc, and $\rho_s=2.5\times10^6 M_\odot$~kpc$^{-3}$
(a normal concentration halo with $M_{\rm vir}\sim1.8\times10^{12}M_\odot$)..
The dotted curves show a high-concentration halo with $\nu=2.79$,
$r_s=22$~kpc, $\rho_s=5.1\times10^6 M_\odot$~kpc$^{-3}$,
and $M_{\rm vir}\sim1.3\times10^{12}M_\odot$;
and a low-concentration halo with $\nu=3.48$,
$r_s=49$~kpc, $\rho_s=1.2\times10^6 M_\odot$~kpc$^{-3}$,
and $M_{\rm vir}\sim2.5\times10^{12}M_\odot$.
}\label{fig:mwdisp}
\end{figure*}

The Milky Way modeling differs slightly from the previous two galaxies by treating 3-D rather than projected profiles (of density and velocity dispersion).  For the mass model, a $\Lambda$CDM motivated halo is included as before, along with stellar disk and bulge components based on models ``I'' and ``II'' from \citet{binney_galactic_2008}, which bracket a plausible range of parameters for these components.  We also tried the disk and bulge model of \citet{widrow_dynamical_2008}, which is roughly halfway between the I and II models of \citet{binney_galactic_2008}.  As shown in the upper left panel of Fig.~\ref{fig:mwdisp}, since these components influence the central regions only, all choices give the same results for $\nu$.  The model I of \citet{binney_galactic_2008} was adopted.

The modeling results are illustrated in Figure~\ref{fig:mwdisp}.  After excluding the region inside $r=$~2~kpc from the density profile fit because of the presence of
a core \citep{parmentier_origin_2005,bica_globular_2006}, a good fit to both the density and radial velocity dispersion profiles is found.  The best-fit value for the MPGC subhalo rarity is \mwnu, where the uncertainty is again driven by the scatter in the assumed halo mass-concentration relation.  The parameters for the DM halos in these models are listed in the Figure caption.  The different DM model derived here is the main reason why a higher $\nu$ is found compared to the earlier work by \citet{moore_globular_2006}.

Note that there is a gap in the MPGC velocity data around $r\sim$~50~kpc.  If the MPGC velocity dispersion in this region were assumed to be as low as for the halo stars (as indicated by the dashed curve in the Figure, from \citealt{gnedin_mass_2010}) then the virial mass and $\nu$ value would need to be lower.  The default halo mass profile is actually similar to that of \citet{gnedin_mass_2010} despite the higher velocity dispersion found here. This is because the high typical anisotropy $\beta \sim 0.65$ in the current model elevates the radial dispersion relative to the $\beta \sim 0.4$ that the former authors assumed, and which would be appropriate for the dark matter particles on average, not for the ``biased'' subset of halo particles that should correspond to the stars and MPGCs.

\subsection{Mass modelling summary}

In the Milky Way and M87, the MPGC kinematics comfortably fit within the \citet{diemand_distribution_2005} theoretical framework and are consistent with other mass tracers.  Though the mass models unfortunately suffer from rather large systematic uncertainties from the $\Lambda$CDM prior, the quality of the models mean the $\nu\sigma$ measurements described below are on firm ground.  The situation for NGC~1407, from a mass-modelling perspective, is less clear.  There appears to be lingering tension between the models and observations, which merit further study.

\subsection{Constraining the reionization epoch}\label{reion}

\begin{figure}
\center
\includegraphics[scale=0.45]{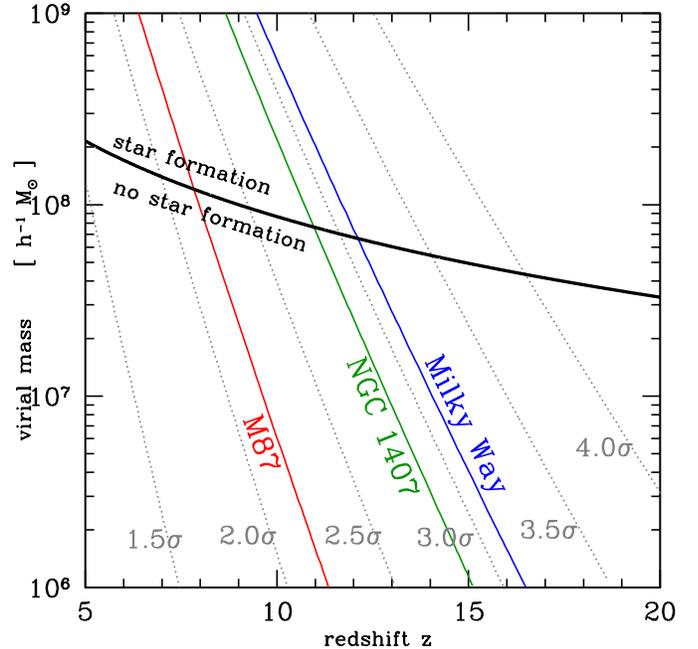}
\caption[Redshift-mass pairs for a given subhalo rarity.]{Evolution of subhalo rarity ($\nu\sigma$) over a range of virial masses and redshifts $z$.  Various $\nu\sigma$ peaks are indicated by dotted, grey lines.  Solid black line shows the mass-redshift pairs, where halos have virial temperatures T$_{\rm vir}=10^4$ K and efficient star formation can take place \citep[curve from ][]{moore_globular_2006}.  The constraints for the galaxies studied here are given as the coloured, solid lines: \mesnus, \nftosnus, and \mwnus~from left to right (M87, NGC~1407 and the Milky Way, respectively).}\label{figmassredshift}
\endcenter
\end{figure}

\begin{table}
 \centering
  \caption[Estimates for $z_{reion}$]{Rarity ($\nu\sigma$) of MPGC progenitor subhalos, the reionization redshifts and minimum virial mass of the progenitors for the target galaxies.}\label{tabreion} 
  \begin{tabular}{@{}llrrr@{}}
\hline
        & environmental &         &             & min. virial \\
        & density       & $\nu$   & $z_{reion}$ & mass \\
        &               &         &             & [$h^{-1}$ \msun] \\
  \hline \hline
Messier 87 & high/cluster & \mesnub   & \meszplain     & $1\times10^8$ \\
NGC 1407   & medium/group & \nftosnub & \nftoszplain   & $8\times10^7$ \\
Milky Way  & low/field    & \mwnub    & \mwzplain  & $7\times10^7$ \\
\hline
mean       &              & \allnub   & \allzplain   & $8\times10^7$ \\
\hline
\end{tabular}
\end{table}

To convert the $\nu\sigma$ constraints from the preceding sections into estimates for the truncation redshift of MPGC formation (e.g. $z_{reion}$), a degeneracy must be broken between subhalo virial mass and redshift for a given $\nu$.  The evolving properties of a given $\nu\sigma$ subhalo can be analytically computed using the \citep{press_formation_1974} Gaussian field formalism.  The value $\nu$ is explicitly defined as:  $\nu=\delta_c / [D(z) \sigma(M_{\rm vir})]$, where the constant $\delta_c$ is the criterion for spherical collapse, $D(z)$ is the linear growth factor and $\sigma(M_{\rm vir})$ is the typical mass-density fluctuation ``height'' for a subhalo of a given virial mass $M_{\rm vir}$ \citep[see e.g. ][]{mo_abundance_2002}.  For the present work, a top hat filter is used to select the characteristic spatial scale.  The CAMB \citep{lewis_efficient_2000} online interface\footnote{{\tt http://lambda.gsfc.nasa.gov/toolbox/tb\_camb\_form.cfm}} was used to generate the linear power spectrum for a WMAP7 cosmology \citep{komatsu_seven-year_2011}.  Figure~\ref{figmassredshift} shows the relationship between the rarity, $\nu$, of a mass-density fluctuation for a given $M_{\rm vir}-$redshift pair.  

Figure~\ref{figmassredshift} also shows how a $\nu\sigma$ measurement alone will not constrain the redshift when MPGC formation is suppressed. Further information is required to break this degeneracy.  Following \citet{moore_globular_2006}, it is assumed that star and MPGC formation can only occur in subhalos that have virial temperatures greater than T$_{\rm vir}=10^4$K, where gas is able to rapidly cool and form stars \citep[e.g. ][]{ostriker_reheating_1996}.  The region in Fig.~\ref{figmassredshift} below the T$_{\rm vir}=10^4$K curve can therefore be excluded.  The region above the T$_{\rm vir}=10^4$K curve can be excluded as well, since MPGC formation would have taken place in subhalos that would ultimately have less or more concentrated spatial distributions than what is observed in the galaxies.  For example, for the Milky Way constraint of $\nu=3.1$, at redshift of $z=10$, subhalos with mass $M_{\rm vir}\sim10^8 M_{\odot}$ could still have formed MPGCs efficiently, which would have led to a spatial distribution resembling $2.5\sigma$ subhalos, contradicting the observations.  At redshifts higher than the intersection between the T$_{\rm vir}=10^4$K relation and the $\nu\sigma$ curve from MPGC observations, only rarer, e.g. $\nu>3.1$, subhalos could form MPGCs, thus the spatial distribution would have been more concentrated than is observed.  The intersection between the $\nu\sigma$ constraint and the T$_{\rm vir}=10^4$K curve is therefore taken to be the MPGC truncation redshift or the local $z_{reion}$ of the host galaxy.

The MPGC $\nu\sigma$ fits and the corresponding reionization redshifts are given in Table~\ref{tabreion} for each galaxy.  The weighted mean of the three $\nu_{MPGC}$ measurements translates into a reionization redshift of \allz.  These constraints are discussed in Section~\ref{discussion} below.

\section{Discussion}\label{discussion}

\begin{figure}
\center
\includegraphics[scale=0.45]{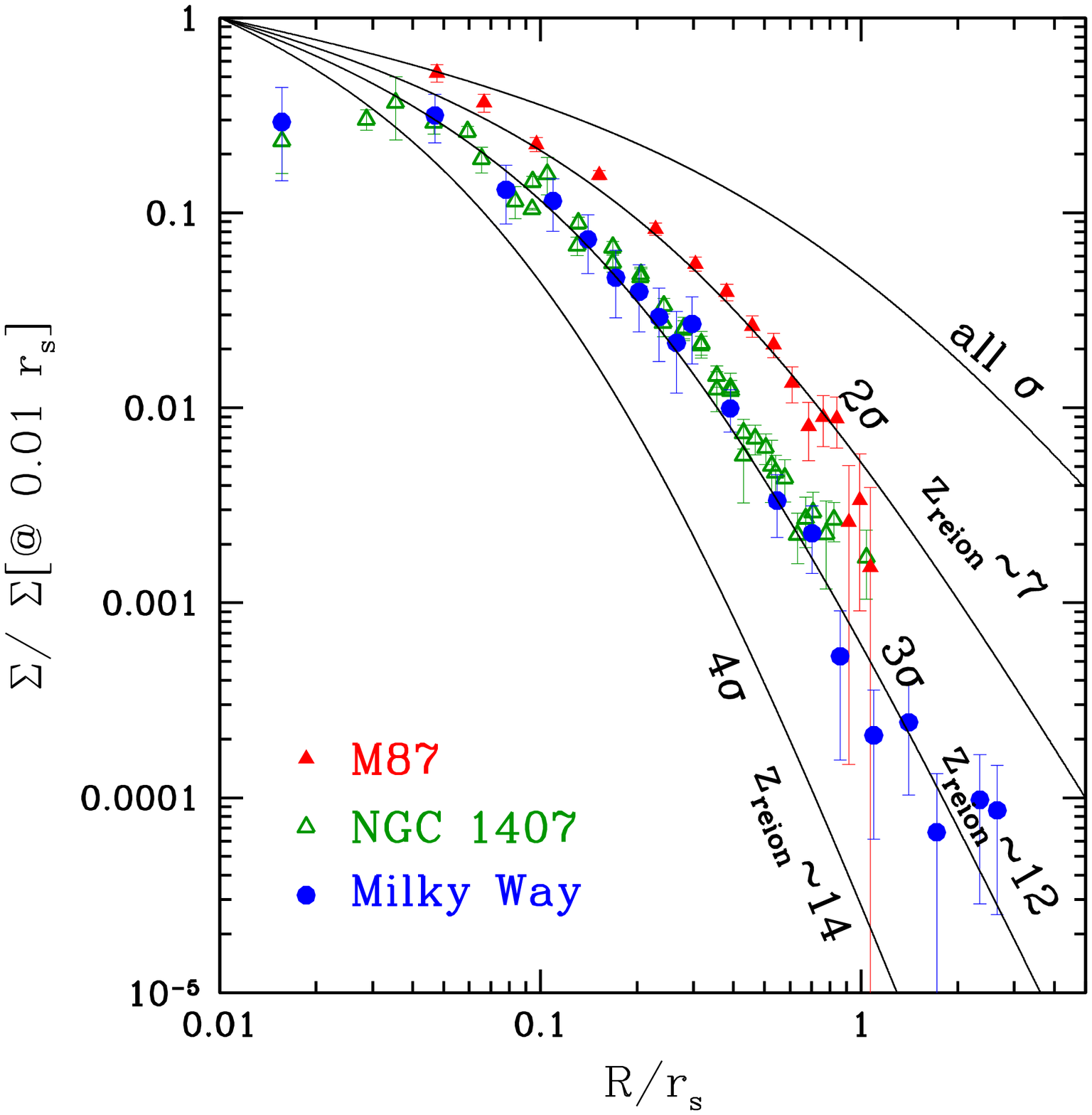}
\caption[Spatial distribution of metal-poor globular clusters.]{Normalized, projected surface density profiles of the metal-poor globular clusters in each of the target galaxies, which are labelled in the figure.  Curves show the subhalo spatial distribution for a range of rarity values (e.g. $\nu\ge2,3,4$ standard deviations from the cosmic mean mass-density), which directly constrain the epoch at which metal-poor globular cluster formation was halted by the local reionization epoch \citep{diemand_distribution_2005,moore_globular_2006}.  From fitting the theoretical expectations for the spatial and kinematic distributions of metal-poor globular clusters (see \S\ref{method}), the epoch of reionization is constrained to be \mwz, \nftosz, and \mesz for the Milky Way, NGC~1407 and M87, respectively.}\label{summary_radial}
\endcenter
\end{figure}

\begin{figure*}
\center
\includegraphics[scale=0.6]{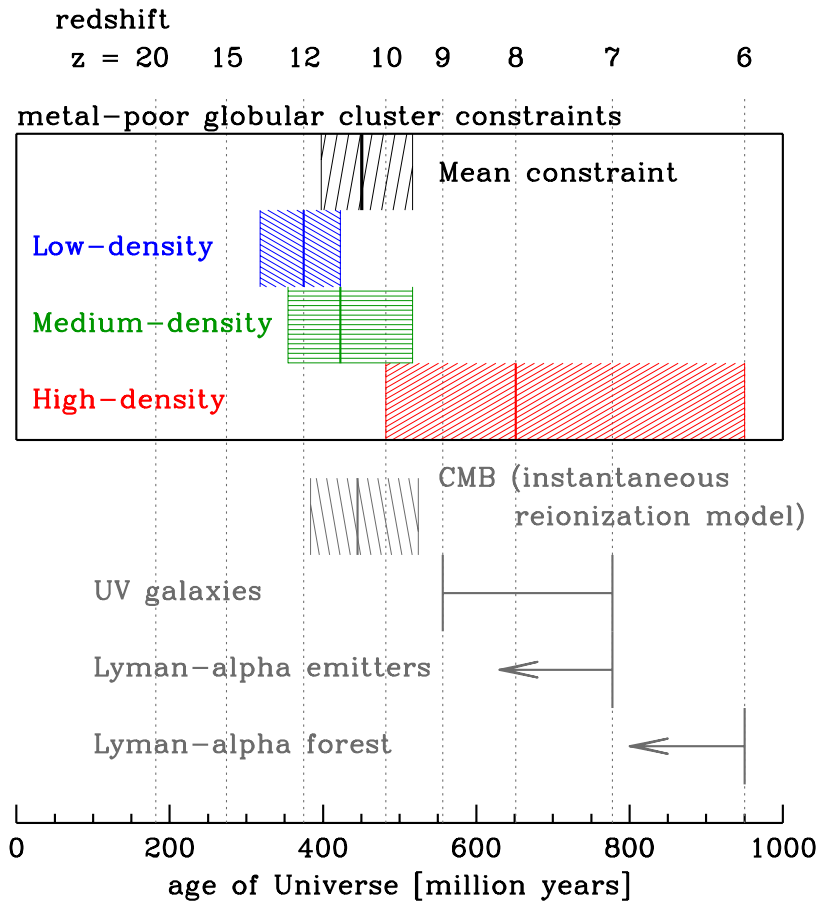}
\caption[Reionization redshift constraints.]{Reionization constraints from the current work and the literature.  The reionization epoch measurements for the local metal-poor globular cluster observations (coloured constraints) are labeled with their local environments (the Milky Way, NGC~1407 and M87 for low-, medium- and high-density environments, respectively) and dominated by uncertainties on galaxy's mass model.  High-redshift constraints include:  cosmic microwave background result from \citet{komatsu_seven-year_2011}, constraints from UV galaxy observations from \citet{robertson_early_2010}, Lyman-$\alpha$ emitters from \citet{kashikawa_completing_2011,pentericci_spectroscopic_2011}, and Lyman-$\alpha$ forest from \citet{fan_constraining_2006}. Arrows indicate an estimated limit on the reionization epoch, while each hatched region contains the 68\% uncertainty about a measurement.}\label{summary_epoch}
\endcenter
\end{figure*}

The reionization redshift ($z_{reion}$) constraints from metal-poor globular cluster observations (see Fig.~\ref{summary_radial}) are summarized in Fig.~\ref{summary_epoch} with existing estimates and limits from the literature. The local constraints from metal-poor globular cluster (MPGC) observations are tabulated in Tab.~\ref{tabreion}.  

The only existing constraints on $z_{reion}$ from MPGCs are for the Milky Way.  As described in \S\ref{intro}, our estimate agrees with that derived in \citet{moore_globular_2006}, who used the same technique employed here.  Also, the high-resolution dark matter Milky Way simulation described in \citet{griffen_globular_2010} can only reproduce the observed spatial distribution of MPGCs and their numbers only when MPGC formation was suppressed by reionization at $z_{reion}\sim13$.  This value agrees with our own estimate within its 1$\sigma$ uncertainty.

\subsection{Environmental propagation of reionization}

The MPGC $z_{reion}$ values and their uncertainties span the redshift range covered by $z_{reion}$ estimates from the literature, with the weighted mean MPGC constraint showing best agreement with the cosmic microwave background (CMB) value and an instantaneous reionization model \citep{larson_seven-year_2011,komatsu_seven-year_2011}.  It is apparent that the M87 $z_{reion}$ measurement is somewhat of an outlier compared to the other MPGC estimates.  Since M87 is located in the high-density environment of a galaxy cluster, it is possible that its lower $z_{reion}$ results from an environmentally-dependent reionization epoch.  Unfortunately, the relatively large systematic uncertainties on the galaxy halo mass models mean the difference between $z_{reion}$ in the highest and lowest density environments (M87 vs. Milky Way $z_{reion}$) is only significant at the $1.7\sigma$ level.

In addition to a correlation between the MPGC spatial distribution and the local reionization epoch, \citet{moore_globular_2006} predicted that $z_{reion}$ should also correlate with the virial mass-normalized MPGC numbers. This is because galaxies with a lower $z_{reion}$ had more time to produce MPGCs.  Both \citet{spitler_connection_2008} and \citet{spitler_new_2009} found no evidence for a correlation between MPGC virial mass specific frequencies and environment.  However, their virial mass estimates were derived from statistical relationships between stellar and virial masses, which are not ideal for use on individual galaxies.

Using the galaxy virial masses derived in Section~\ref{method} and MPGC GC numbers ($N_{MPGC}$) from Section~\ref{observations}, the following MPGC virial mass specific frequencies \citep[$V_{MPGC}$ ][]{spitler_connection_2008} are found:  $V_{MPGC} = N_{MPGC}/(M_{\rm vir}/10^{12}) = 27.5^{+12.9}_{-8.4}, 6.7^{+2.7}_{-1.9}, 6.1^{+2.4}_{-1.7}$, for M87, NGC~1407, and the Milky Way, respectively.  The errors are again dominated by systematic uncertainties in the galaxy mass models.  M87 has a $V_{MPGC}$ value that is larger than the Milky Way or NGC~1407, at the $2.5\sigma$ significance level.  The normalized MPGC numbers thus lend further support to the idea that reionization finished in M87 at lower redshifts compared to the other galaxies.  In general, massive galaxies at the centres of galaxy clusters (e.g. M87) are known to have enhanced relative number of GCs when normalized to the galaxy's star light \citep[e.g. ][]{harris_globular_1991,mclaughlin_efficiency_1999}.  Whether this holds when the MPGC numbers are instead normalized to virial masses (which should not be directly influenced by reionization) will need to be explored in a future work.

The above results provide an exciting hint that reionization completed first in low-density environments.

\subsection{The theoretical context}

A number of large-scale radiative transfer simulations have been run to understand the propagation of reionization through different galaxy environments.  From analysis of these simulations, it is found that proto-galaxies in high-density environments produce significant numbers of ionizing sources early on and thus are the first locations to reionize \citep[][however see \citealt{weinmann_dependence_2007}]{iliev_simulating_2006,finlator_late_2009,iliev_reionization_2011}. 

Such simulations typically do not have a large enough dynamical range to capture both the environmental propagation of reionization over large volumes and the small-scale physics that are important for understanding localized affects.  For example, when an inhomogeneous intergalactic medium is incorporated into reionization models, dense pockets of neutral gas can survive even after surrounding regions had completed the reionization process \citep{miralda-escude_reionization_2000,ciardi_simulating_2003,furlanetto_taxing_2005,iliev_effect_2008,choudhury_inside-out_2009,mitra_reionization_2011,mitra_joint_2012}.  These pockets tend to be located in dense environments, where recombination rates can stay high despite their proximity to large reservoirs of ionizing sources.  This means the intergalactic medium in dense environments started to reionize first, but only completely reionized {\it after} lower density environments had finished.

The trend hinted at by the metal-poor globular cluster observations agrees with this latter theoretical work.  Under this scenario, MPGCs in a high-density environment continued forming to lower redshifts in self-shielded pockets, even as low environmental-density regions were being completely ionized.  The progenitor molecular clouds of MPGCs located in more vulnerable environments (like that surrounding the Milky Way) were not surrounded by gas with sufficiently high recombination rates to survive the ionizing front that originated within or passed through the region.

Finally, the range of the MPGC constraints on $z_{reion}$ agrees with the range of literature constraints (see Fig.~\ref{summary_epoch}).  This seems to support models of reionization that are globally extended in duration \citep[e.g. ][]{mitra_reionization_2011,mitra_joint_2012}. 

\section{Conclusions}

The spatial and kinematic properties of metal-poor globular clusters (MPGCs) provide an exciting hint that reionization was inhomogeneous.  In the local Universe, reionization completed first in low-density environments around $z_{reion}\sim11-12$ (for the Milky Way and large central group elliptical NGC~1407) and finished last in the high-density environment around the Virgo cluster at $z_{reion}\sim8$.  While uncertainties on the galaxy halo mass models limit the strength of this conclusion, another property, the relative number of metal-poor globular clusters, supports this interpretation.  The apparent environmental dependence of $z_{reion}$ also aligns with theoretical work that finds that the ionization front propagated from high- to low-density environments, with the high densities finishing reionization last.  Furthermore, the apparent agreement between the range of $z_{reion}$ constraints from MPGC observations and those from the literature may provide additional support to models where reionization was globally prolonged in duration.

More observational work, especially to improve the individual galaxy halo mass models, is required to confirm the results found here.  It will also be important to model MPGC formation in large-scale, radiative transfer simulations with realistic models of star cluster dynamical evolution.  This will help refine the technique utilized here and clarify the relationship between reionization and the truncation of MPGC formation.  For example, given that high star formation rates and gas densities were needed to produce such dense star clusters, it is possible that MPGCs trace {\it the end} of reionization, where ionizing radiation was able to finally penetrate the densest gas clouds.  

Expanding this initial study to other galaxies will yield a reionization map of the local Universe, further aiding our understanding of galaxy formation and evolution over cosmic time.

If reionization is not responsible for the truncation of MPGC formation, the results presented here provide unique constraints on the formation histories of MPGCs.   Furthermore, since MPGCs and stellar halos of the Milky Way share spatial and chemical properties \citep[][ see also M31, \citealt{huxor_exploring_2011}]{helmi_stellar_2008,martell_building_2011} they might share a common origin.  If this is the case for other galaxies, then MPGC observations can be used to constrain the star formation history of their stellar halos.

Finally, the power of the Diemand-Moore technique is that it does not care about the detailed star formation and enrichment physics -- it only requires an observable tracer object that accreted onto the galaxy at early times \citep{diemand_distribution_2005}.

\vspace{-0.7cm} \section*{Acknowledgements}

We want to acknowledge the useful comments provided by the anonymous reviewer.  We also thank Chris Blake for assistance with the cosmological derivations and Anna Sippel for help with the initial Subaru reductions. LS was supported by the ARC Discovery Programme grants DP0770233 and DP1094370. This work was supported by the National Science Foundation through grants AST-0808099, AST-0909237, and AST-1109878.  DF thanks the ARC Discovery Programme for support. This paper was based in part on data collected at Subaru Telescope, which is operated by the National Astronomical Observatory of Japan.  Some of the Subaru data were acquired with the timeswap Gemini program GN-2006B-C-18. Some of the data presented herein were obtained at the W.~M.~Keck Observatory, which is operated as a scientific partnership among the California Institute of Technology, the University of California and the National Aeronautics and Space Administration. The Observatory was made possible by the generous financial support of the W.~M.~Keck Foundation. Observations reported here were obtained at the MMT Observatory, a joint facility of the Smithsonian Institution and the University of Arizona.

\vspace{-0.7cm} \bibliographystyle{mn2e}
\bibliography{sci}

\begin{thebibliography}{134}
\expandafter\ifx\csname natexlab\endcsname\relax\def\natexlab#1{#1}\fi

\bibitem[{Ashman \& Zepf(1992)}]{ashman_formation_1992}
Ashman K.~M., Zepf S.~E., 1992, ApJ, 384, 50

\bibitem[{Baek {et~al}\mbox{.}(2010)Baek, Semelin, Di~Matteo, Revaz, \&
  Combes}]{baek_reionization_2010}
Baek S., Semelin B., Di~Matteo P., Revaz Y., Combes F., 2010, A\&A, 523, 4

\bibitem[{Barkana \& Loeb(2001)}]{barkana_beginning:_2001}
Barkana R., Loeb A., 2001, Physics Reports, 349, 125

\bibitem[{Battaglia {et~al}\mbox{.}(2006)Battaglia, Helmi, Morrison, Harding,
  Olszewski, Mateo, Freeman, Norris, \& Shectman}]{battaglia_erratum:_2006}
Battaglia G. {et~al.}, 2006, MNRAS, 370, 1055

\bibitem[{Beasley {et~al}\mbox{.}(2002)Beasley, Baugh, Forbes, Sharples, \&
  Frenk}]{beasley_formation_2002}
Beasley M.~A., Baugh C.~M., Forbes D.~A., Sharples R.~M., Frenk C.~S., 2002,
  MNRAS, 333, 383

\bibitem[{Bekki(2005)}]{bekki_galactic_2005}
Bekki K., 2005, ApJ, 626, L93

\bibitem[{Bekki {et~al}\mbox{.}(2008)Bekki, Yahagi, Nagashima, \&
  Forbes}]{bekki_origin_2008}
Bekki K., Yahagi H., Nagashima M., Forbes D.~A., 2008, MNRAS, 387, 1131

\bibitem[{Bica {et~al}\mbox{.}(2006)Bica, Bonatto, Barbuy, \&
  Ortolani}]{bica_globular_2006}
Bica E., Bonatto C., Barbuy B., Ortolani S., 2006, A\&A, 450, 105

\bibitem[{Binggeli {et~al}\mbox{.}(1987)Binggeli, Tammann, \&
  Sandage}]{binggeli_studies_1987}
Binggeli B., Tammann G.~A., Sandage A., 1987, The AJ, 94, 251

\bibitem[{Binney \& Tremaine(2008)}]{binney_galactic_2008}
Binney J., Tremaine S., 2008, Galactic Dynamics: Second Edition

\bibitem[{Boley {et~al}\mbox{.}(2009)Boley, Lake, Read, \&
  Teyssier}]{boley_globular_2009}
Boley A.~C., Lake G., Read J., Teyssier R., 2009, ApJ Letters, 706, L192

\bibitem[{Bolton {et~al}\mbox{.}(2010)Bolton, Becker, Wyithe, Haehnelt, \&
  Sargent}]{bolton_first_2010}
Bolton J.~S., Becker G.~D., Wyithe J. S.~B., Haehnelt M.~G., Sargent W. L.~W.,
  2010, MNRAS, 406, 612

\bibitem[{Boulade {et~al}\mbox{.}(2003)Boulade, Charlot, Abbon, Aune, Borgeaud,
  Carton, Carty, {Da Costa}, Deschamps, Desforge, Eppell\'e, Gallais, Gosset,
  Granelli, Gros, de~Kat, Loiseau, Ritou, Rouss\'e, Starzynski, Vignal, \&
  Vigroux}]{boulade_megacam:_2003}
Boulade O. {et~al.}, 2003, in , pp. 72--81

\bibitem[{Bouwens {et~al}\mbox{.}(2011)Bouwens, Illingworth, Oesch, Trenti,
  Labbe, Franx, Stiavelli, Carollo, van Dokkum, \&
  Magee}]{bouwens_lower-luminosity_2011}
Bouwens R.~J. {et~al.}, 2011, arXiv:1105.2038B

\bibitem[{Bouwens {et~al}\mbox{.}(2010)Bouwens, Illingworth, Oesch, Trenti,
  Stiavelli, Carollo, Franx, van Dokkum, Labb\'e, \& Magee}]{bouwens_very_2010}
Bouwens R.~J. {et~al.}, 2010, ApJ, 708, L69

\bibitem[{Bowman \& Rogers(2010)}]{bowman_lower_2010}
Bowman J.~D., Rogers A. E.~E., 2010, Nature, 468, 796

\bibitem[{Brodie {et~al}\mbox{.}(2011)Brodie, Romanowsky, Strader, \&
  Forbes}]{brodie_relationships_2011}
Brodie J.~P., Romanowsky A.~J., Strader J., Forbes D.~A., 2011, 1109.5696

\bibitem[{Brodie \& Strader(2006)}]{brodie_extragalactic_2006}
Brodie J.~P., Strader J., 2006, ARAA, 44, 193

\bibitem[{Bromm \& Clarke(2002)}]{bromm_formation_2002}
Bromm V., Clarke C.~J., 2002, ApJ, 566, L1

\bibitem[{Brough {et~al}\mbox{.}(2006)Brough, Forbes, Kilborn, Couch, \&
  Colless}]{brough_eridanus_2006}
Brough S., Forbes D.~A., Kilborn V.~A., Couch W., Colless M., 2006, MNRAS, 369,
  1351

\bibitem[{Bunker {et~al}\mbox{.}(2010)Bunker, Wilkins, Ellis, Stark, Lorenzoni,
  Chiu, Lacy, Jarvis, \& Hickey}]{bunker_contribution_2010}
Bunker A.~J. {et~al.}, 2010, MNRAS, 409, 855

\bibitem[{Choudhury {et~al}\mbox{.}(2009)Choudhury, Haehnelt, \&
  Regan}]{choudhury_inside-out_2009}
Choudhury T.~R., Haehnelt M.~G., Regan J., 2009, MNRAS, 394, 960

\bibitem[{Ciardi {et~al}\mbox{.}(2003)Ciardi, Stoehr, \&
  White}]{ciardi_simulating_2003}
Ciardi B., Stoehr F., White S. D.~M., 2003, MNRAS, 343, 1101

\bibitem[{Cole \& Kaiser(1989)}]{cole_biased_1989}
Cole S., Kaiser N., 1989, MNRAS, 237, 1127

\bibitem[{C\^ot\'e {et~al}\mbox{.}(1998)C\^ot\'e, Marzke, \&
  West}]{ct_formation_1998}
C\^ot\'e P., Marzke R.~O., West M.~J., 1998, ApJ, 501, 554

\bibitem[{Das {et~al}\mbox{.}(2010)Das, Gerhard, Churazov, \&
  Zhuravleva}]{das_steepening_2010}
Das P., Gerhard O., Churazov E., Zhuravleva I., 2010, MNRAS, 409, 1362

\bibitem[{Diemand {et~al}\mbox{.}(2005)Diemand, Madau, \&
  Moore}]{diemand_distribution_2005}
Diemand J., Madau P., Moore B., 2005, MNRAS, 364, 367

\bibitem[{Dijkstra {et~al}\mbox{.}(2004)Dijkstra, Haiman, \&
  Loeb}]{dijkstra_limit_2004}
Dijkstra M., Haiman Z., Loeb A., 2004, ApJ, 613, 646

\bibitem[{Doherty {et~al}\mbox{.}(2009)Doherty, Arnaboldi, Das, Gerhard,
  Aguerri, Ciardullo, Feldmeier, Freeman, Jacoby, \&
  Murante}]{doherty_edge_2009}
Doherty M. {et~al.}, 2009, A\&A, 502, 771

\bibitem[{Dopita {et~al}\mbox{.}(2011)Dopita, Krauss, Sutherland, Kobayashi, \&
  Lineweaver}]{dopita_re-ionizing_2011}
Dopita M.~A., Krauss L.~M., Sutherland R.~S., Kobayashi C., Lineweaver C.~H.,
  2011, Astrophysics and Space Science, 335, 345

\bibitem[{Duffy {et~al}\mbox{.}(2008)Duffy, Schaye, Kay, \&
  Dalla~Vecchia}]{duffy_dark_2008}
Duffy A.~R., Schaye J., Kay S.~T., Dalla~Vecchia C., 2008, MNRAS, 390, L64

\bibitem[{Dutton {et~al}\mbox{.}(2011)Dutton, Conroy, van~den Bosch, Simard,
  Mendel, Courteau, Dekel, More, \& Prada}]{dutton_dark_2011}
Dutton A.~A. {et~al.}, 2011, MNRAS, 416, 322

\bibitem[{Erb {et~al}\mbox{.}(2006)Erb, Steidel, Shapley, Pettini, Reddy, \&
  Adelberger}]{erb_stellar_2006}
Erb D.~K., Steidel C.~C., Shapley A.~E., Pettini M., Reddy N.~A., Adelberger
  K.~L., 2006, ApJ, 646, 107

\bibitem[{Faber {et~al}\mbox{.}(2003)Faber, Phillips, Kibrick, Alcott, Allen,
  Burrous, Cantrall, Clarke, \& et~al.}]{faber_deimos_2003}
Faber S.~M. {et~al.}, 2003, in , pp. 1657--1669

\bibitem[{Fabricant {et~al}\mbox{.}(2005)Fabricant, Fata, Roll, Hertz,
  Caldwell, Gauron, Geary, {McLeod}, \& et~al.}]{fabricant_hectospec_2005}
Fabricant D. {et~al.}, 2005, PASP, 117, 1411

\bibitem[{Fall \& Zhang(2001)}]{fall_dynamical_2001}
Fall S.~M., Zhang Q., 2001, ApJ, 561, 751

\bibitem[{Fan {et~al}\mbox{.}(2006)Fan, Strauss, Becker, White, Gunn, Knapp,
  Richards, Schneider, Brinkmann, \& Fukugita}]{fan_constraining_2006}
Fan X. {et~al.}, 2006, The AJ, 132, 117

\bibitem[{Finkelstein {et~al}\mbox{.}(2010)Finkelstein, Papovich, Giavalisco,
  Reddy, Ferguson, Koekemoer, \& Dickinson}]{finkelstein_stellar_2010}
Finkelstein S.~L., Papovich C., Giavalisco M., Reddy N.~A., Ferguson H.~C.,
  Koekemoer A.~M., Dickinson M., 2010, ApJ, 719, 1250

\bibitem[{Finlator {et~al}\mbox{.}(2009)Finlator, Özel, Dav\'e, \&
  Oppenheimer}]{finlator_late_2009}
Finlator K., Özel F., Dav\'e R., Oppenheimer B.~D., 2009, MNRAS, 400, 1049

\bibitem[{Forbes {et~al}\mbox{.}(2011)Forbes, Spitler, Strader, Romanowsky,
  Brodie, \& Foster}]{forbes_evidence_2011}
Forbes D., Spitler L., Strader J., Romanowsky A., Brodie J., Foster C., 2011,
  1101.3575

\bibitem[{Forbes {et~al}\mbox{.}(1997)Forbes, Brodie, \&
  Grillmair}]{forbes_origin_1997}
Forbes D.~A., Brodie J.~P., Grillmair C.~J., 1997, AJ, 113, 1652

\bibitem[{Forbes {et~al}\mbox{.}(2006)Forbes, {S\'anchez-Bl\'azquez}, Phan,
  Brodie, Strader, \& Spitler}]{forbes_imaging_2006}
Forbes D.~A., {S\'anchez-Bl\'azquez} P., Phan A. T.~T., Brodie J.~P., Strader
  J., Spitler L., 2006, MNRAS, 366, 1230

\bibitem[{Foster {et~al}\mbox{.}(2010)Foster, Forbes, Proctor, Strader, Brodie,
  \& Spitler}]{foster_deriving_2010}
Foster C., Forbes D.~A., Proctor R.~N., Strader J., Brodie J.~P., Spitler
  L.~R., 2010, The AJ, 139, 1566

\bibitem[{Furlanetto \& Oh(2005)}]{furlanetto_taxing_2005}
Furlanetto S.~R., Oh S.~P., 2005, MNRAS, 363, 1031

\bibitem[{Gebhardt \& {Kissler-Patig}(1999)}]{gebhardt_globular_1999}
Gebhardt K., {Kissler-Patig} M., 1999, AJ, 118, 1526

\bibitem[{Gnedin {et~al}\mbox{.}(2010)Gnedin, Brown, Geller, \&
  Kenyon}]{gnedin_mass_2010}
Gnedin O.~Y., Brown W.~R., Geller M.~J., Kenyon S.~J., 2010, ApJ, 720, L108

\bibitem[{Gray \& Scannapieco(2010)}]{gray_formation_2010}
Gray W.~J., Scannapieco E., 2010, ApJ, 718, 417

\bibitem[{Griffen {et~al}\mbox{.}(2010)Griffen, Drinkwater, Thomas, Helly, \&
  Pimbblet}]{griffen_globular_2010}
Griffen B.~F., Drinkwater M.~J., Thomas P.~A., Helly J.~C., Pimbblet K.~A.,
  2010, MNRAS, 405, 375

\bibitem[{Hansen {et~al}\mbox{.}(2007)Hansen, Anderson, Brewer, Dotter,
  Fahlman, Hurley, Kalirai, King, Reitzel, Richer, Rich, Shara, \&
  Stetson}]{hansen_white_2007}
Hansen B. M.~S. {et~al.}, 2007, ApJ, 671, 380

\bibitem[{Hansen {et~al}\mbox{.}(2004)Hansen, Richer, Fahlman, Stetson, Brewer,
  Currie, Gibson, Ibata, Rich, \& Shara}]{hansen_hubble_2004}
Hansen B. M.~S. {et~al.}, 2004, ApJSS, 155, 551

\bibitem[{Harris(1991)}]{harris_globular_1991}
Harris W.~E., 1991, ARAA, 29, 543

\bibitem[{Harris(2010)}]{harris_new_2010}
Harris W.~E., 2010, 1012.3224

\bibitem[{Harris {et~al}\mbox{.}(2006)Harris, Whitmore, Karakla, Okon, Baum,
  Hanes, \& Kavelaars}]{harris_globular_2006}
Harris W.~E., Whitmore B.~C., Karakla D., Okon W., Baum W.~A., Hanes D.~A.,
  Kavelaars J.~J., 2006, ApJ, 636, 90

\bibitem[{Hasegawa {et~al}\mbox{.}(2009)Hasegawa, Umemura, \&
  Kitayama}]{hasegawa_formation_2009}
Hasegawa K., Umemura M., Kitayama T., 2009, MNRAS, 397, 1338

\bibitem[{Helmi(2008)}]{helmi_stellar_2008}
Helmi A., 2008, A\&A Review, 15, 145

\bibitem[{Huxor {et~al}\mbox{.}(2011)Huxor, Ferguson, Tanvir, Irwin, Mackey,
  Ibata, Bridges, Chapman, \& Lewis}]{huxor_exploring_2011}
Huxor A.~P. {et~al.}, 2011, MNRAS, 414, 770

\bibitem[{Iliev {et~al}\mbox{.}(2006)Iliev, Mellema, Pen, Merz, Shapiro, \&
  Alvarez}]{iliev_simulating_2006}
Iliev I.~T., Mellema G., Pen U., Merz H., Shapiro P.~R., Alvarez M.~A., 2006,
  MNRAS, 369, 1625

\bibitem[{Iliev {et~al}\mbox{.}(2011)Iliev, Moore, Gottl\"ober, Yepes, Hoffman,
  \& Mellema}]{iliev_reionization_2011}
Iliev I.~T., Moore B., Gottl\"ober S., Yepes G., Hoffman Y., Mellema G., 2011,
  MNRAS, 413, 2093

\bibitem[{Iliev {et~al}\mbox{.}(2008)Iliev, Shapiro, {McDonald}, Mellema, \&
  Pen}]{iliev_effect_2008}
Iliev I.~T., Shapiro P.~R., {McDonald} P., Mellema G., Pen U., 2008, MNRAS,
  391, 63

\bibitem[{Kashikawa {et~al}\mbox{.}(2011)Kashikawa, Shimasaku, Matsuda, Egami,
  Jiang, Nagao, Ouchi, Malkan, Hattori, Ota, Taniguchi, Okamura, Ly, Iye,
  Furusawa, Shioya, Shibuya, Ishizaki, \&
  Toshikawa}]{kashikawa_completing_2011}
Kashikawa N. {et~al.}, 2011, ApJ, 734, 119

\bibitem[{Komatsu {et~al}\mbox{.}(2011)Komatsu, Smith, Dunkley, Bennett, Gold,
  Hinshaw, Jarosik, Larson, Nolta, Page, Spergel, Halpern, Hill, Kogut, Limon,
  Meyer, Odegard, Tucker, Weiland, Wollack, \&
  Wright}]{komatsu_seven-year_2011}
Komatsu E. {et~al.}, 2011, ApJSS, 192, 18

\bibitem[{Krauss \& Chaboyer(2003)}]{krauss_age_2003}
Krauss L.~M., Chaboyer B., 2003, Science, 299, 65

\bibitem[{Kundu \& Whitmore(2001)}]{kundu_new_2001}
Kundu A., Whitmore B.~C., 2001, AJ, 122, 1251

\bibitem[{Kundu \& Zepf(2007)}]{kundu_bimodal_2007}
Kundu A., Zepf S.~E., 2007, ApJ, 660, L109

\bibitem[{Labb\'e {et~al}\mbox{.}(2010)Labb\'e, Gonz\'alez, Bouwens,
  Illingworth, Franx, Trenti, Oesch, van Dokkum, Stiavelli, Carollo, Kriek, \&
  Magee}]{labbe_star_2010}
Labb\'e I. {et~al.}, 2010, ApJ, 716, L103

\bibitem[{Lacey \& Cole(1993)}]{lacey_merger_1993}
Lacey C., Cole S., 1993, MNRAS, 262, 627

\bibitem[{Larsen {et~al}\mbox{.}(2001)Larsen, Brodie, Huchra, Forbes, \&
  Grillmair}]{larsen_properties_2001}
Larsen S.~S., Brodie J.~P., Huchra J.~P., Forbes D.~A., Grillmair C.~J., 2001,
  AJ, 121, 2974

\bibitem[{Larson {et~al}\mbox{.}(2011)Larson, Dunkley, Hinshaw, Komatsu, Nolta,
  Bennett, Gold, Halpern, Hill, Jarosik, Kogut, Limon, Meyer, Odegard, Page,
  Smith, Spergel, Tucker, Weiland, Wollack, \& Wright}]{larson_seven-year_2011}
Larson D. {et~al.}, 2011, ApJSS, 192, 16

\bibitem[{Lewis {et~al}\mbox{.}(2000)Lewis, Challinor, \&
  Lasenby}]{lewis_efficient_2000}
Lewis A., Challinor A., Lasenby A., 2000, ApJ, 538, 473

\bibitem[{Lidz {et~al}\mbox{.}(2007)Lidz, {McQuinn}, Zaldarriaga, Hernquist, \&
  Dutta}]{lidz_quasar_2007}
Lidz A., {McQuinn} M., Zaldarriaga M., Hernquist L., Dutta S., 2007, ApJ, 670,
  39

\bibitem[{Lorenzoni {et~al}\mbox{.}(2011)Lorenzoni, Bunker, Wilkins, Stanway,
  Jarvis, \& Caruana}]{lorenzoni_star-forming_2011}
Lorenzoni S., Bunker A.~J., Wilkins S.~M., Stanway E.~R., Jarvis M.~J., Caruana
  J., 2011, MNRAS, 414, 1455

\bibitem[{Macci{\`o} {et~al}\mbox{.}(2008)Macci{\`o}, Dutton, \& van~den
  Bosch}]{maccio_concentration_2008}
Macci{\`o} A.~V., Dutton A.~A., van~den Bosch F.~C., 2008, MNRAS, 391, 1940

\bibitem[{Mamon \& {\L}okas(2005)}]{mamon_dark_2005}
Mamon G.~A., {\L}okas E.~L., 2005, MNRAS, 363, 705

\bibitem[{Mannucci {et~al}\mbox{.}(2009)Mannucci, Cresci, Maiolino, Marconi,
  Pastorini, Pozzetti, Gnerucci, Risaliti, Schneider, Lehnert, \&
  Salvati}]{mannucci_lsd:_2009}
Mannucci F. {et~al.}, 2009, MNRAS, 398, 1915

\bibitem[{Martell {et~al}\mbox{.}(2011)Martell, Smolinski, Beers, \&
  Grebel}]{martell_building_2011}
Martell S.~L., Smolinski J.~P., Beers T.~C., Grebel E.~K., 2011, A\&A, 534, 136

\bibitem[{{McLaughlin}(1999{\natexlab{a}})}]{mclaughlin_efficiency_1999}
{McLaughlin} D.~E., 1999{\natexlab{a}}, AJ, 117, 2398

\bibitem[{{McLaughlin}(1999{\natexlab{b}})}]{mclaughlin_evidence_1999}
{McLaughlin} D.~E., 1999{\natexlab{b}}, ApJ, 512, L9

\bibitem[{{Miralda-Escud\'e} {et~al}\mbox{.}(2000){Miralda-Escud\'e}, Haehnelt,
  \& Rees}]{miralda-escude_reionization_2000}
{Miralda-Escud\'e} J., Haehnelt M., Rees M.~J., 2000, ApJ, 530, 1

\bibitem[{Mitra {et~al}\mbox{.}(2011)Mitra, Choudhury, \&
  Ferrara}]{mitra_reionization_2011}
Mitra S., Choudhury T.~R., Ferrara A., 2011, MNRAS, 413, 1569

\bibitem[{Mitra {et~al}\mbox{.}(2012)Mitra, Choudhury, \&
  Ferrara}]{mitra_joint_2012}
Mitra S., Choudhury T.~R., Ferrara A., 2012, MNRAS, 419, 1480

\bibitem[{Miyazaki {et~al}\mbox{.}(2002)Miyazaki, Komiyama, Sekiguchi, Okamura,
  Doi, Furusawa, Hamabe, Imi, Kimura, Nakata, Okada, Ouchi, Shimasaku, Yagi, \&
  Yasuda}]{miyazaki_subaru_2002}
Miyazaki S. {et~al.}, 2002, PASJ, 54, 833

\bibitem[{Mo \& White(2002)}]{mo_abundance_2002}
Mo H.~J., White S. D.~M., 2002, MNRAS, 336, 112

\bibitem[{Moore {et~al}\mbox{.}(2006)Moore, Diemand, Madau, Zemp, \&
  Stadel}]{moore_globular_2006}
Moore B., Diemand J., Madau P., Zemp M., Stadel J., 2006, MNRAS, 368, 563

\bibitem[{Moore {et~al}\mbox{.}(1998)Moore, Governato, Quinn, Stadel, \&
  Lake}]{moore_resolving_1998}
Moore B., Governato F., Quinn T., Stadel J., Lake G., 1998, ApJ, 499, L5

\bibitem[{Muratov \& Gnedin(2010)}]{muratov_modeling_2010}
Muratov A.~L., Gnedin O.~Y., 2010, ApJ, 718, 1266

\bibitem[{Murphy {et~al}\mbox{.}(2011)Murphy, Gebhardt, \&
  Adams}]{murphy_galaxy_2011}
Murphy J.~D., Gebhardt K., Adams J.~J., 2011, ApJ, 729, 129

\bibitem[{Napolitano {et~al}\mbox{.}(2009)Napolitano, Romanowsky, Coccato,
  Capaccioli, Douglas, Noordermeer, Gerhard, Arnaboldi, de~Lorenzi, Kuijken,
  Merrifield, {O'Sullivan}, Cortesi, Das, \&
  Freeman}]{napolitano_planetary_2009}
Napolitano N.~R. {et~al.}, 2009, MNRAS, 393, 329

\bibitem[{Navarro {et~al}\mbox{.}(1996)Navarro, Frenk, \&
  White}]{navarro_structure_1996}
Navarro J.~F., Frenk C.~S., White S. D.~M., 1996, ApJ, 462, 563

\bibitem[{Neilsen \& Tsvetanov(1999)}]{neilsen_color_1999}
Neilsen E.~H., Tsvetanov Z.~I., 1999, ApJ Letters, 515, L13

\bibitem[{Ocvirk \& Aubert(2011)}]{ocvirk_signature_2011}
Ocvirk P., Aubert D., 2011, 1108.1193

\bibitem[{Oke {et~al}\mbox{.}(1995)Oke, Cohen, Carr, Cromer, Dingizian, Harris,
  Labrecque, Lucinio, Schaal, Epps, \& Miller}]{oke_keck_1995}
Oke J.~B. {et~al.}, 1995, PASP, 107, 375

\bibitem[{Ostriker \& Gnedin(1996)}]{ostriker_reheating_1996}
Ostriker J.~P., Gnedin N.~Y., 1996, ApJ, 472, L63

\bibitem[{Ouchi {et~al}\mbox{.}(2010)Ouchi, Shimasaku, Furusawa, Saito,
  Yoshida, Akiyama, Ono, Yamada, Ota, Kashikawa, Iye, Kodama, Okamura, Simpson,
  \& Yoshida}]{ouchi_statistics_2010}
Ouchi M. {et~al.}, 2010, ApJ, 723, 869

\bibitem[{Parmentier \& Grebel(2005)}]{parmentier_origin_2005}
Parmentier G., Grebel E.~K., 2005, MNRAS, 359, 615

\bibitem[{Peng {et~al}\mbox{.}(2006)Peng, Jord\'an, C\^ot\'e, Blakeslee,
  Ferrarese, Mei, West, Merritt, Milosavljevic, \& Tonry}]{peng_acs_2006}
Peng E.~W. {et~al.}, 2006, ApJ, 639, 95

\bibitem[{Pentericci {et~al}\mbox{.}(2011)Pentericci, Fontana, Vanzella,
  Castellano, Grazian, Dijkstra, Boutsia, Cristiani, Dickinson, Giallongo,
  Giavalisco, Maiolino, Moorwood, Paris, \&
  Santini}]{pentericci_spectroscopic_2011}
Pentericci L. {et~al.}, 2011, ApJ, 743, 132

\bibitem[{Pipino {et~al}\mbox{.}(2007)Pipino, Puzia, \&
  Matteucci}]{pipino_formation_2007}
Pipino A., Puzia T.~H., Matteucci F., 2007, ApJ, 665, 295

\bibitem[{Pota {et~al}\mbox{.}(2011)Pota, Forbes, Brodie, Romanowsky, \&
  et}]{pota_survey_2011}
Pota V., Forbes D.~A., Brodie J.~P., Romanowsky A.~J., et a., 2011, {MNRAS} in
  prep.

\bibitem[{Power {et~al}\mbox{.}(2009)Power, Wynn, Combet, \&
  Wilkinson}]{power_primordial_2009}
Power C., Wynn G.~A., Combet C., Wilkinson M.~I., 2009, MNRAS, 395, 1146

\bibitem[{Prada {et~al}\mbox{.}(2011)Prada, Klypin, Cuesta, {Betancort-Rijo},
  \& Primack}]{prada_halo_2011}
Prada F., Klypin A.~A., Cuesta A.~J., {Betancort-Rijo} J.~E., Primack J., 2011,
  arXiv:1104.5130

\bibitem[{Press \& Schechter(1974)}]{press_formation_1974}
Press W.~H., Schechter P., 1974, ApJ, 187, 425

\bibitem[{Proctor {et~al}\mbox{.}(2009)Proctor, Forbes, Romanowsky, Brodie,
  Strader, Spolaor, Mendel, \& Spitler}]{proctor_probing_2009}
Proctor R.~N., Forbes D.~A., Romanowsky A.~J., Brodie J.~P., Strader J.,
  Spolaor M., Mendel J.~T., Spitler L., 2009, MNRAS, 398, 91

\bibitem[{Rhode {et~al}\mbox{.}(2010)Rhode, Windschitl, \&
  Young}]{rhode_wiyn_2010}
Rhode K.~L., Windschitl J.~L., Young M.~D., 2010, The AJ, 140, 430

\bibitem[{Rhode {et~al}\mbox{.}(2007)Rhode, Zepf, Kundu, \&
  Larner}]{rhode_global_2007}
Rhode K.~L., Zepf S.~E., Kundu A., Larner A.~N., 2007, The AJ, 134, 1403

\bibitem[{Rhode {et~al}\mbox{.}(2005)Rhode, Zepf, \&
  Santos}]{rhode_metal-poor_2005}
Rhode K.~L., Zepf S.~E., Santos M.~R., 2005, ApJ, 630, L21

\bibitem[{Robertson {et~al}\mbox{.}(2010)Robertson, Ellis, Dunlop, {McLure}, \&
  Stark}]{robertson_early_2010}
Robertson B.~E., Ellis R.~S., Dunlop J.~S., {McLure} R.~J., Stark D.~P., 2010,
  Nature, 468, 49

\bibitem[{Romanowsky {et~al}\mbox{.}(2011)Romanowsky, Strader, Brodie, \&
  et}]{romanowsky_ongoing_2011}
Romanowsky A.~J., Strader J., Brodie J.~P., et a., 2011, {ApJ}, submitted

\bibitem[{Romanowsky {et~al}\mbox{.}(2009)Romanowsky, Strader, Spitler,
  Johnson, Brodie, Forbes, \& Ponman}]{romanowsky_mappingdark_2009}
Romanowsky A.~J., Strader J., Spitler L.~R., Johnson R., Brodie J.~P., Forbes
  D.~A., Ponman T., 2009, AJ, 137, 4956

\bibitem[{Salvaterra {et~al}\mbox{.}(2011)Salvaterra, Ferrara, \&
  Dayal}]{salvaterra_simulating_2011}
Salvaterra R., Ferrara A., Dayal P., 2011, MNRAS, 414, 847

\bibitem[{Santos(2003)}]{santos_high-redshift_2003}
Santos M.~R., 2003, p. 348

\bibitem[{Scannapieco {et~al}\mbox{.}(2004)Scannapieco, Weisheit, \&
  Harlow}]{scannapieco_triggering_2004}
Scannapieco E., Weisheit J., Harlow F., 2004, ApJ, 615, 29

\bibitem[{Shapiro {et~al}\mbox{.}(2010)Shapiro, Genzel, \&
  F\"orster~Schreiber}]{shapiro_star-forming_2010}
Shapiro K.~L., Genzel R., F\"orster~Schreiber N.~M., 2010, MNRAS, 403, L36

\bibitem[{Shen \& Gebhardt(2010)}]{shen_supermassive_2010}
Shen J., Gebhardt K., 2010, ApJ, 711, 484

\bibitem[{Sheth \& Tormen(1999)}]{sheth_large-scale_1999}
Sheth R.~K., Tormen G., 1999, MNRAS, 308, 119

\bibitem[{Spitler(2010)}]{spitler_first_2010}
Spitler L.~R., 2010, MNRAS, 406, 1125

\bibitem[{Spitler \& Forbes(2009)}]{spitler_new_2009}
Spitler L.~R., Forbes D.~A., 2009, MNRAS, 392, L1

\bibitem[{Spitler {et~al}\mbox{.}(2008{\natexlab{a}})Spitler, Forbes, \&
  Beasley}]{spitler_extendingbaseline:_2008}
Spitler L.~R., Forbes D.~A., Beasley M.~A., 2008{\natexlab{a}}, MNRAS, 389,
  1150

\bibitem[{Spitler {et~al}\mbox{.}(2008{\natexlab{b}})Spitler, Forbes, Strader,
  Brodie, \& Gallagher}]{spitler_connection_2008}
Spitler L.~R., Forbes D.~A., Strader J., Brodie J.~P., Gallagher J.~S.,
  2008{\natexlab{b}}, MNRAS, 385, 361

\bibitem[{Srbinovsky \& Wyithe(2010)}]{srbinovsky_fraction_2010}
Srbinovsky J.~A., Wyithe J. S.~B., 2010, PASA, 27, 110

\bibitem[{Strader {et~al}\mbox{.}(2007)Strader, Beasley, \&
  Brodie}]{strader_globular_2007}
Strader J., Beasley M.~A., Brodie J.~P., 2007, AJ, 133, 2015

\bibitem[{Strader {et~al}\mbox{.}(2006)Strader, Brodie, Spitler, \&
  Beasley}]{strader_globular_2006}
Strader J., Brodie J.~P., Spitler L., Beasley M.~A., 2006, AJ, 132, 2333

\bibitem[{Strader {et~al}\mbox{.}(2011)Strader, Romanowsky, Brodie, Spitler,
  Beasley, Arnold, Tamura, Sharples, \& Arimoto}]{strader_wide-field_2011}
Strader J. {et~al.}, 2011, ApJSS, 197, 33

\bibitem[{{Trujillo-Gomez} {et~al}\mbox{.}(2011){Trujillo-Gomez}, Klypin,
  Primack, \& Romanowsky}]{trujillo-gomez_galaxies_2011}
{Trujillo-Gomez} S., Klypin A., Primack J., Romanowsky A.~J., 2011, ApJ, 742,
  16

\bibitem[{van~den Bergh(2001)}]{van_den_bergh_how_2001}
van~den Bergh S., 2001, ApJ, 559, L113

\bibitem[{Vesperini {et~al}\mbox{.}(2003)Vesperini, Zepf, Kundu, \&
  Ashman}]{vesperini_modeling_2003}
Vesperini E., Zepf S.~E., Kundu A., Ashman K.~M., 2003, ApJ, 593, 760

\bibitem[{Weinmann {et~al}\mbox{.}(2007)Weinmann, Macci{\`o}, Iliev, Mellema,
  \& Moore}]{weinmann_dependence_2007}
Weinmann S.~M., Macci{\`o} A.~V., Iliev I.~T., Mellema G., Moore B., 2007,
  MNRAS, 381, 367

\bibitem[{White \& Springel(2000)}]{white_where_2000}
White S. D.~M., Springel V., 2000, p. 327

\bibitem[{Widrow {et~al}\mbox{.}(2008)Widrow, Pym, \&
  Dubinski}]{widrow_dynamical_2008}
Widrow L.~M., Pym B., Dubinski J., 2008, ApJ, 679, 1239

\bibitem[{Willott {et~al}\mbox{.}(2010)Willott, Delorme, Reyl\'e, Albert,
  Bergeron, Crampton, Delfosse, Forveille, Hutchings, {McLure}, Omont, \&
  Schade}]{willott_canada-france_2010}
Willott C.~J. {et~al.}, 2010, The AJ, 139, 906

\bibitem[{Wise \& Abel(2008)}]{wise_how_2008}
Wise J.~H., Abel T., 2008, ApJ, 684, 1

\bibitem[{Wolf {et~al}\mbox{.}(2010)Wolf, Martinez, Bullock, Kaplinghat, Geha,
  Muñoz, Simon, \& Avedo}]{wolf_accurate_2010}
Wolf J., Martinez G.~D., Bullock J.~S., Kaplinghat M., Geha M., Muñoz R.~R.,
  Simon J.~D., Avedo F.~F., 2010, MNRAS, 406, 1220

\bibitem[{Xue {et~al}\mbox{.}(2008)Xue, Rix, Zhao, Re~Fiorentin, Naab,
  Steinmetz, van~den Bosch, Beers, Lee, Bell, Rockosi, Yanny, Newberg, Wilhelm,
  Kang, Smith, \& Schneider}]{xue_milky_2008}
Xue X.~X. {et~al.}, 2008, ApJ, 684, 1143

\bibitem[{Yan {et~al}\mbox{.}(2010)Yan, Windhorst, Hathi, Cohen, Ryan,
  {O'Connell}, \& {McCarthy}}]{yan_galaxy_2010}
Yan H., Windhorst R.~A., Hathi N.~P., Cohen S.~H., Ryan R.~E., {O'Connell}
  R.~W., {McCarthy} P.~J., 2010, Research in A\&A, 10, 867

\bibitem[{Zinn(1985)}]{zinn_globular_1985}
Zinn R., 1985, ApJ, 293, 424

\end{thebibliography}

\label{lastpage}

\end{document}